\documentclass[sigconf]{acmart}
\usepackage{xcolor}
\usepackage{soul}
\usepackage{graphicx}
\usepackage{caption}
\usepackage{balance}
\AtBeginDocument{%
  \providecommand\BibTeX{{%
    \normalfont B\kern-0.5em{\scshape i\kern-0.25em b}\kern-0.8em\TeX}}}

\copyrightyear{2023} 
\acmYear{2023} 
\setcopyright{rightsretained} 
\acmConference[CHI '23]{Proceedings of the 2023 CHI Conference on Human Factors in Computing Systems}{April 23--28, 2023}{Hamburg, Germany}
\acmBooktitle{Proceedings of the 2023 CHI Conference on Human Factors in Computing Systems (CHI '23), April 23--28, 2023, Hamburg, Germany}\acmDOI{10.1145/3544548.3581175}
\acmISBN{978-1-4503-9421-5/23/04}

\begin{document}

\title[The Entoptic Field Camera]{The Entoptic Field Camera as Metaphor-Driven Research-through-Design with AI Technologies}


\author{Jesse Josua Benjamin}
\orcid{0000-0003-3391-3060}
\authornote{Second affiliation with Imagination, Lancaster University, Lancaster, United Kingdom.}
\affiliation{%
  \mbox{\institution{University of Twente}
  \country{Netherlands}}
}
\email{j.j.benjamin@utwente.nl}

\author{Heidi Biggs}
\affiliation{
    \institution{The Pennsylvania State University}
    \country{USA}
}

\author{Arne Berger}
\affiliation{
    \institution{Anhalt University of Applied Sciences}
    \country{Germany}
}

\author{Julija Rukanskait\.e}
\affiliation{
    \institution{Independent Researcher}
    \country{Sweden}
}

\author{Michael Heidt}
\affiliation{
    \institution{Anhalt University of Applied Sciences}
    \country{Germany}
}

\author{Nick Merrill}
\affiliation{
    \institution{University of California, Berkeley}
    \country{USA}
}

\author{James Pierce}
\affiliation{
    \institution{University of Washington}
    \country{USA}
}

\author{Joseph Lindley}
\affiliation{
    \institution{Lancaster University}
    \country{United Kingdom}
}

\renewcommand{\shortauthors}{Benjamin, et al.}
\begin{abstract}
 Artificial intelligence (AI) technologies are widely deployed in smartphone photography; and prompt-based image synthesis models have rapidly become commonplace. In this paper, we describe a Research-through-Design (RtD) project which explores this shift in the means and modes of image production via the creation and use of the Entoptic Field Camera. Entoptic phenomena usually refer to perceptions of floaters or bright blue dots stemming from the physiological interplay of the eye and brain. We use the term entoptic as a metaphor to investigate how the material interplay of data and models in AI technologies shapes human experiences of reality. Through our case study using first-person design and a field study, we offer implications for critical, reflective, more-than-human and ludic design to engage AI technologies; the conceptualisation of an RtD research space which contributes to AI literacy discourses; and outline a research trajectory concerning materiality and design affordances of AI technologies.
\end{abstract}

\ccsdesc[500]{Human-centered computing~Human computer interaction (HCI)}
\ccsdesc[300]{Human-centered computing~Interaction design theory, concepts and paradigms}
\ccsdesc[300]{Human-centered computing~Field studies}
\ccsdesc[300]{Computing methodologies~Artificial intelligence}
\keywords{research through design, artificial intelligence, materiality, GAN, image synthesis, technological mediation}

\begin{teaserfigure}
  \centering
  \includegraphics[width=0.99\textwidth]{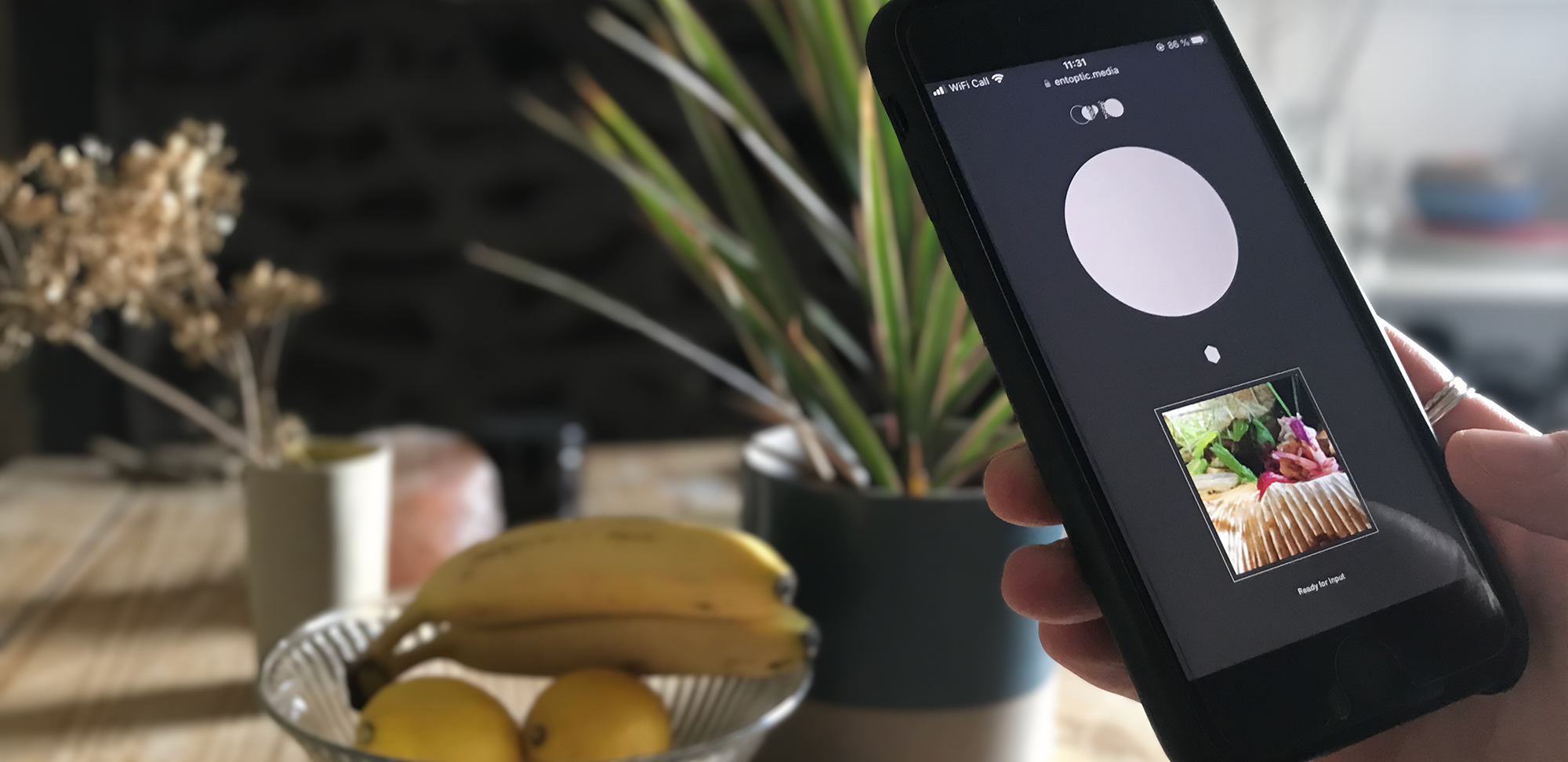}
  \caption{The Entoptic Field Camera is designed to facilitate in situ use of GAN techniques to create synthetic images based on user input images, thus prompting reflections on how AI technologies shape the experience of particular realities and the design opportunities this implies.}
  \captionsetup{justification=centering,margin=2cm}
  \Description{Photograph of the Entoptic Field Camera prototype being used in situ on a smartphone (foreground) to create a GAN-generated image of items on a table (background).}
  \label{fig:teaser}
\end{teaserfigure}

\maketitle

\section{Introduction}
In the summer of 2022, the outputs of image synthesis models (e.g., OpenAI's Dall-E, TikTok’s AI Greenscreen, Stability AI’s stable diffusion, Midjourney, or Google's Imagen) that generate images from text prompts are everywhere on social media. They are fast becoming a means of expression for professional artists and general enthusiasts,\footnote{\url{https://twitter.com/images_ai}, accessed 09/12/2022.} a shortcut for companies in image production (see e.g.~\cite{warzel_where_2022}) and a basis for memes.\footnote{\url{https://twitter.com/weirddalle}, accessed 09/12/2022.} As AI artists Herndon and Dryhurst note, AI image technology---like photography before it---is rapidly transitioning ``from a scientific to an expressive medium''~\cite{herndon_infinite_2022}. With unprecedented levels of accessibility, such AI technologies are now leaving scientific or corporate settings and becoming deployed for everyday purposes such as creative expression (see e.g.~\cite{vincent_engine_2022}); while also being commodified as, for instance, marketplaces for prompts\footnote{See e.g. \url{https://promptbase.com}, accessed 09/12/2022.} or end-user image editing solutions (see e.g.~\cite{senzer_photobombs_2021}). Especially prompt-based image synthesis models suggest to some that AI technologies are finally offering utility to everyone (see e.g.~\cite{jiang_promptmaker_2022}). At the same time, this very utilization occurs against a backdrop of phenomena that imply more fundamental questions regarding the relationship of experience, materiality and design of AI technologies that remain to be tackled. 

Attending to such phenomena is important, since AI image technologies also serve as a litmus-test for what AI technologies in general are seemingly capable of.\footnote{It is no coincidence that in parallel to the flood of prompt-based image synthesis, a cross-disciplinary academic and popular discourse emerged surrounding a Google research engineer’s belief in a specific AI technology’s sentience---and subsequent firing [see e.g.~\cite{bogost_googles_2022,bratton_model_2022}.} One example is particularly illustrative of this ongoing need. In the summer of 2020, California was plagued by extensive wildfires, resulting in skies being colored in outlandishly deep red tones. However, people found that taking smartphone images resulted in tones that `defaulted' to gray and blue; breaking the relationship between a local phenomenon and its representation. As Bogost noted, the smartphone cameras never `expected' the deep red skies~\cite{bogost_your_2020} due to the training data of embedded AI technologies. While it is not news that smartphones employ AI technologies to overcome the physical constraints of lenses and sensors~\cite{chen_reason_2019}, this example shows that the reality-shaping influence of AI technologies can run in parallel, and simultaneously `below,' an explicit utilization such as improving images. Even below the `threshold of utility' for a particular AI technology to be deployed in or as a product, therefore, there are ways that AI technologies render realities and make them accessible to human experience. For many users, this may have been the first time they were presented with a visible clue that computational, artificially-intelligent photographic technologies were at work behind the scenes whenever they snapped a photo with their smartphone. Like images and imaging instruments always have~\cite{mitchell_picture_1995,kittler_optical_2010} we therefore argue that AI image technologies imply not only aesthetic, but rather deeply epistemological concerns. Our concern is that current enthusiasm over synthetic imaging capacities may overshadow these subtle, but equally profound ways that AI technologies shape how variant realities are experienced by people. 
Drawing from post-phenomenology~\cite{ihde_technology_1990}, and mediation theory in particular~\cite{verbeek_what_2006}, we therefore ask: \textit{What specific opportunities for design does the shaping of experiencing particular realities by AI technologies hold?}

In this paper, we engage this question by way of a metaphor-driven Research-through-Design (RtD) project. In doing so, we treat AI technologies as a design \textit{material} for exploration and experimentation, rather than as means or ends in themselves---an approach which the HCI community has long pursued regarding computing and emerging information technologies (see e.g. ~\cite{lim_discovery-driven_2013,sundstrom_inspirational_2011}). Indeed, RtD (and design research) with AI technologies is a highly active research field within the HCI community~\cite{benjamin_machine_2021,biggs_high_2020,wakkary_morse_2017,lindley_researching_2020} and regarding the use of metaphors especially~\cite{dove_monsters_2020,murray-rust_metaphors_2022}. In our approach, we engage AI image technologies by way of a principal design metaphor: \textit{entoptic phenomena}. Entoptic (Greek: `within vision') phenomena are perceptions that do not directly correspond to material reality, but rather stem from the physiological interplay of eye and brain (cf.~\cite{noauthor_entoptic_2010,helmholtz_handbuch_1867,lewis-williams_signs_1988}). A common example for an entoptic phenomenon are `floaters.' These are small dark shapes, spots, or web-like structures that appear in perception. They are caused by material inside the eye that physically interferes with the light as it passes through the lens and vitreous fluid on its route to hitting the retina,  subsequently being rendered as particular perceptual shapes by the brain. 

Used as a design metaphor, entoptic phenomena support the conceptualization of AI technologies as generating particular perceptions through the interplay of their material components (i.e., models, algorithms, data). The metaphor was further inspired by previous arguments in ~\cite{benjamin_machine_2021}, who propose the concept of ``pattern leakage'' to show how AI technologies may shape the world they are intended to represent due to the use of probabilistic techniques. Following the entoptic metaphor, we conceptualized and designed the \textit{Entoptic Field Camera}.\footnote{\url{https://entoptic.media/cam/}, accessed 07/09/2022. Note that this prototype is the result of a later iteration, in which one functionality (the Manual mode, see section 3.2.1.) was removed. See also \url{https://twitter.com/entoptic__media}, accessed 09/15/2022.} In effect, the Entoptic Field Camera is a web application which generates a synthetic output image in response to a user's input image; and is specifically built for situated use with a smartphone (see Figure ~\ref{fig:teaser}). By analogy, it therefore embodies subtle cases of everyday reality shaping by AI technologies, such as in the wildfire sky example noted above. We document and discuss the development of the metaphor, the design of the Entoptic Field Camera as well as a field study in which the authors engaged this prototype reflectively. We found that the metaphor as well as its concretization in a prototype brought forth a series of implications for designing with and through AI technologies; and outline four distinct contributions:

\begin{enumerate}
    \item A case study of the conceptualization, development, design and deployment of an RtD artefact---the Entoptic Field Camera---that enables situated and reflective engagement with AI technologies along a central design metaphor;
    \item Novel design implications that make AI technologies available to ludic, critical, reflective, and more-than-human design methodologies;
    \item The conceptualization of a research space for RtD that can contribute to discourses around AI literacy by engaging AI materiality;
    \item A sketch for a research trajectory that further investigates the relationship of AI technologies, 21st century design and its affordances.
\end{enumerate}

\section{Background}
In this section, we first reflect on the roles of images and imaging in AI research and as applied in consumer smartphones. Then, we show how design research and particularly RtD in HCI has extensively thematized and made use of images and imaging. Lastly, we find that this general capacity has not yet significantly been applied to AI technologies, but note that metaphor-driven approaches hold promise for further work.

\subsection{Images and Imaging in AI Research and Application}
Images and imaging play an important role for the design, development and adaptation of artificial intelligence (AI) technologies (e.g., neural networks) in terms of application contexts, socio-cultural imagination as well as communication of engineering achievements.
In the technical fields, the prowess of AI technologies is frequently demonstrated using tasks from image recognition and generation—even if the tasks that AI technologies are designed for are unrelated to imaging per se. The technology company OpenAI, for instance, first showcased technical advances in Natural Language Processing by generating images from text prompts\footnote{\url{https://openai.com/blog/image-gpt/}, accessed 16/11/2021.}; preceding the latter explosion of ``diffusion''~\cite{ho_denoising_2020} models into popular discourse and cultural expression mentioned in the introduction. In technical HCI work, fields such as explainable AI (XAI), Fairness Accountability and Transparency (FAccT) or interpretable machine learning (ML) deploy algorithmic techniques which `image' information extracted from more complex AI technology pipelines. Such approaches often visualize features or generate saliency maps (see e.g.~\cite{hohman_gamut:_2019,kinkeldey_towards_2019,mordvintsev_inceptionism:_2015,olah_building_2018}) with the goal of explanatory guidance. In scientific applications, AI technologies are not just the target of but also means for imaging. For instance, the first-ever image of the M87* black hole was generated through an AI imaging pipeline that itself drew on AI-generated synthetic data to fill in blindspots of the employed telescopes~\cite{bouman_computational_2016}.  

On a more general level, however, imaging itself is undergoing a change due to AI technologies. 
Periodically, this comes to the surface as in the discussed example of images of Californian wildfire skies. Such extreme cases evidence a \textit{modus operandi}---smartphone cameras are now AI-enabled, overcoming the physical limits of their small-sized photo lenses by relying on trained models. Indeed, Chen notes how contemporary consumer-grade smartphones employ ``computational photography, which automatically processes images to look more professional''~\cite{chen_reason_2019}; meaning introducing AI technologies which can, for instance, simulate a wide aperture lens’ bokeh, handle low-light situations, or zoom beyond the material capacities of steel and glass~\cite{qian_bggan_2020}. The specific AI technology that most implementations are based on is the Generative-Adversarial Network (GAN) architecture proposed by Goodfellow and colleagues in 2014; and as noted, more recently, diffusion models such as DALL-E, CLIP, Stable Diffusion or Midjourney have become popular for prompt-based image synthesis. Focussing on GANs, this powerful and indeed controversial\footnote{GANs are highly prominent in current conversations around so-called deepfakes, fully synthetic yet photorealistic images that are easily employed for misinformation, see e.g.~\cite{mirsky_creation_2021}.} AI technology is based on a simple premise: two neural networks competing with each other. A `generator' network is given an input (e.g., an image, text, etc.), and attempts to synthesize it according to learned patterns (e.g., adjusting colors, removing chromatic aberrations or compression artefacts). The result is then fed to a `discriminator' network, which statistically decides whether the generator’s output and the original input can be distinguished, and until this is not the case, the generative model starts a next iteration of synthesis. Goodfellow and colleagues illustrate this concept evocatively by suggesting that ``the generative model can be thought of as analogous to a team of counterfeiters, trying to produce fake currency and use it without detection, while the discriminative model is analogous to the police, trying to detect the counterfeit currency''~\cite{goodfellow_generative_2014}.

\begin{figure}
    \centering
    \includegraphics[width=1\columnwidth]{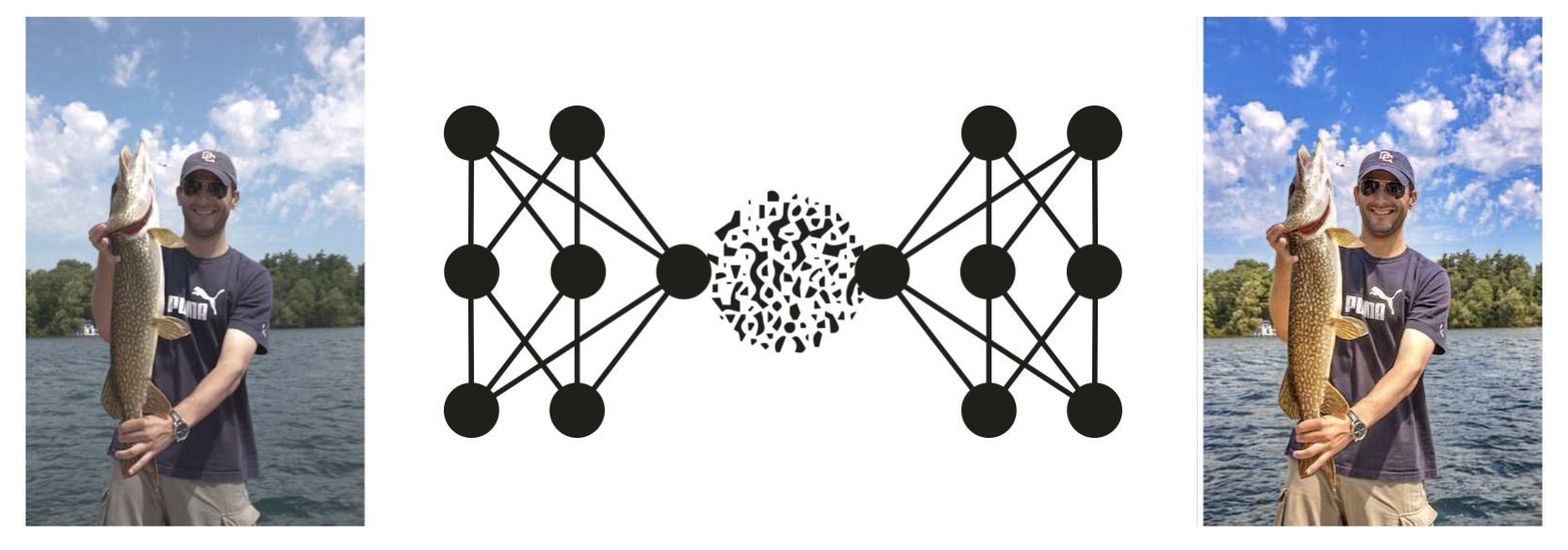}
    \caption{Conceptual diagram for the employment of Generative Adversarial Networks (GANs) in computational photography. An input image (left) is processed by competing neural networks based on learned patterns (center) into an `improved' output image (right). Images from~\cite{chen_deep_2018}.}
    \Description{Conceptual diagram for the employment of Generative Adversarial Networks (GANs) in computational photography. An input image (left, photograph of a man holding a fish against a river, forest and sky in the background) is processed by competing neural networks based on learned patterns (center, abstract representation) into an `improved' output image (right, input image with adapted color grading and simulated high-dynamic range).}
    \label{fig:gan-concept}
\end{figure}

In the context of computational photography, the capacity to generate outputs based on learned features can be used to construct wholly synthetic images that resemble an input, but also a wider range of more subtle corrections such as color grading, removing chromatic aberrations and simulating high-dynamic range (see e.g. ~\cite{chen_deep_2018}, also Figure~\ref{fig:gan-concept}). Aside from the overall visual appearance of images, GAN techniques can also be used to manipulate the content of images. The so-called `inpainting' technique, in which a portion of an image is masked and to be filled in by a GAN (see e.g. ~\cite{yi_contextual_2020}) or similar model, can also already be found in consumer-grade smartphones such as the Google Pixel 6 for the removal of unwanted ``photobombs''~\cite{senzer_photobombs_2021}. In sum, AI technologies are in the process of reshaping images as well as imaging practices. Importantly, the AI-driven correction, inpainting (and a recent development, outpainting) and the whole-sale synthesis of images are carried out not only in the latest prompt-based image synthesis models, but are already at work in everyday technologies. In this regard, they fall within the purview of RtD and design research work that focuses on how situated experiences and practices with technologies develop and what particular opportunities for design unfold. In the next section, we therefore briefly reflect on existing engagement with images and imaging.


\subsection{Images and Imaging in Design Research and RtD}\label{sec:images-rtd}

Images and their production---whether pictures or video---are hugely important to many, if not most, HCI applications. In design research, imaging technologies and especially the relationship between images and the perception of reality have been amply explored. To fully cover this is beyond the scope of this paper, therefore we briefly supply evidence for the rich tradition of engaging images and imaging in design research. To frame this further, as noted in Frayling’s conceptualization of RtD~\cite{frayling_research_1994}, what we seek to demonstrate is that the field’s ability to conduct ``practical experiment[s]'' with technologies holds potential for work in this space. There is a long line of research emphasizing the importance of using visually rich content in HCI design research~\cite{blevis_visual_2012} such as photo-rich research products which showcase design and material processes~\cite{jarvis_attention_2012,lin_tilting_2019,zheng_sensing_2019} as well as reflecting on values, scenarios, and aesthetics of images all at once~\cite{liu_photography_2018,mckinnon_finding_2016}. 

Photography also has been used as a tool in design methods, the most famous example being that of photographs as tools in cultural probe kits~\cite{dunne_cultural_1999}, but photographs have also been used as a way to situated speculation~\cite{desjardins_bespoke_2019}, and reflect on theoretical concepts~\cite{liu_photography_2018}. For example, Blevis’ own image-heavy work often reflects on the complexities and nuances of sustainability and globalization (see e.g.~\cite{blevis_all_2019}), while Desjardins and colleagues imbue their photographic method with feminist theory~\cite{desjardins_bespoke_2019}, and Liu, Bardzell and Bardzell use photography to showcase natureculture as a value for design~\cite{liu_photography_2018}. Concerning non-photographic imaging, Elsden and colleagues’ \textit{Zoom Obscura}~\cite{elsden_zoom_2022} interrogates video-conferencing through counterfunctional designs; for instance by way of speculative AI-generated doppelgangers. As a particularly noteworthy project, Pierce and Paulos's \textit{Obscura 1C Digital Camera}~\cite{pierce_making_2015} used a photography-device and a design-led method to elicit heuristic insights, deliberately prioritising conceptual and imaginary uses of the prototype, over empirically evaluated first-hand uses. 
Standing in for other RtD and HCI work that does what Verbeek refers to as philosophy ``by other means'' (~\cite{verbeek_beyond_2015} see e.g. ~\cite{hauser_deployments_2018,wakkary_morse_2017}), the Obscura 1C is an example for probing the materiality of novel technologies (e.g., digital photography) that have become rapidly adopted; and in turn developing a ``conceptual vocabulary''~\cite{benjamin_machine_2021} that enables further designer inquiries. 

In short, we argue that design research and RtD no doubt has the potential to address latent propositions embodied in objects that draw from AI image technologies---thereby generating conceptual and practical insights for designing or assessing novel experiences, products or scenarios. At the same time, however, AI image technologies have rarely been directly addressed; mirroring the relatively scarce engagement with AI technologies as a design research material. Hence, in the next section we address how RtD and design researchers more generally have thus far framed their approaches to AI technologies.

\subsection{RtD and Metaphoric Approaches to AI}\label{sec:metaphors-rtd}
The address of AI technologies in terms of design in the HCI community has centered primarily on how AI can be made tangible as a design material for designers (see e.g.~\cite{yang_re-examining_2020} on a series of studies). At the same time, there is a growing body of research that centers on the materiality---i.e. the arrangement of components and the materialization of their functional interoperation in specific outputs---of AI technologies to develop conceptual vocabularies, and particularly, metaphors. The use of metaphors by designers is a staple that goes back to canonical texts (e.g.~\cite{schon_generative_1979}). As has been pointed out by numerous authors in HCI and beyond~\cite{agre_computation_1997,agre_toward_1997,blackwell_reification_2006,murray-rust_metaphors_2022}, the term AI in itself is already metaphorical at root in that it assigns, symbolically, capacities to technical devices that are usually reserved for living beings; and especially, human beings. 

While the applicability of particular metaphors (e.g., learning, sentience) has been debated since the early days of AI research~\cite{mcdermott_artificial_1976}, the stubborn opacity of contemporary AI technologies at runtime~\cite{burrell_how_2016,dourish_algorithms_2016} also means that it requires in-depth practical and conceptual engagement to gain a foothold on them. Therefore, it is no surprise that design research work with AI technologies has centered on metaphors that make AI technologies approachable to design practice by suggesting symbolic translations. To date, analogical metaphors in the form of `X-as-Y' are the most common. For instance, Murray-Rust and colleagues propose
``alternative metaphors for designers working with AI metaphors''~\cite{murray-rust_metaphors_2022} such as corporations, geofoam, or fossils. Each of these carries with them an implicit symbolic register of terms, shapes and movements; thereby holding potential design approaches that operate below surface-level concerns such as opacity or technical accuracy. RtD’s capacity for practical experiments is a natural fit for probing emerging technologies for new metaphors (see e.g.~\cite{ricketts_mental_2019}). As an example, Fayard and Dove’s work on the metaphor of ML-as-monster~\cite{dove_monsters_2020} has shown the generative potential of this approach: through metaphor-driven making of `ML-monsters,' the participants in their co-design workshop discerned ``territories of concern at an early stage of design and [pointed] to where exploratory inquiry may be most needed.'' In this regard, RtD is already working on opening up AI technologies for extensive and inquisitive designerly approaches. 

However, we note a significant lack in the metaphorical approaches to AI technologies, which constitutes a gap for further RtD-based inquiry. On a general level, we argue that `linguistic' design research approaches, such as discerning and testing metaphors, may replicate a utilitarian understanding of AI technologies as discrete entities---which, as has been conclusively articulated by Lindh~\cite{lindh_as_2016}, perpetuates existing presuppositions regarding information technologies. In other words, while metaphors may offer a generative lens on what AI technologies \textit{are}, this does not necessarily extend to what AI technologies \textit{do}. Therefore, we propose that metaphors which echo the broad spectrum of phenomena associated with the deployment of AI technologies---such as the California wildfire skies mentioned above---are needed. We frame this argument by drawing on Ihde’s proposal that technologies themselves ought to be considered ``language-analogues''~\cite{ihde_expanding_1998} that make the world `legible' in particular ways. Accordingly, the potential for metaphorical approaches to AI technologies lies not only with understanding more \textit{about} AI, but rather also in experimentally understanding about how a \textit{world} that is shaped by AI technologies is `read.' Lindley and colleagues~\cite{lindley_researching_2020,lindley_signs_2020} have conducted RtD work on legibility with regards to AI technologies; which they deem to be particularly promising regarding ``the emerging reality of living with AI and making those relationships more legible.'' While Lindley and colleagues approach this promise through iconographic experimentation, we echo their sentiment and argue that RtD using metaphors can offer critical openings for design and theory to engage not what an AI technology \textit{is}, but rather what it \textit{does}.

In the following, we show how we have approached this gap in an RtD project consisting of the metaphor-driven conception of an artefact, its implementation and subsequent deployment in a field study. Subsequently, we show how our findings have specific design implications for the RtD and HCI community, offer propositions for the role of RtD with regards to AI literacy, and further reflect on future work regarding the relationship of AI materiality and design.

\section{The Entoptic Field Camera}
In this section, we present an RtD project in which an artefact, the Entoptic Field Camera, was conceptualized, developed, and studied through first-person making and a field study. The Entoptic Field Camera is a web application which, through GAN techniques, generates a synthetic output image in response to a user's input image; specifically built for situated use with a smartphone. By analogy, it therefore embodies subtle cases of reality shaping by AI technologies such as the wildfire sky example noted above. First, we discuss the genesis of the entoptic metaphor. Second, we present our methodological approach, consisting of a first-person design process and a field study with secondary users. Third, we present findings at various levels of emergence.

\subsection{Development of the Entoptic Metaphor}
At root, the entoptic metaphor was conceptualized as an expansion on the conceptual vocabulary developed in ~\cite{benjamin_machine_2021} around the general concept of ``thingly uncertainty,'' and specifically of the notion of \textit{pattern leakage}. Thingly uncertainty was proposed to make the qualitative distinction of AI technologies to other information technologies tangible for designers, namely in that AI-driven artefacts exist in variable relations to their users and environments along a probabilistic continuum. The concept has since been taken up in the HCI community (see e.g. ~\cite{brand_design_2021,scurto_prototyping_2021,rukanskaite_tuning_2021}). In turn, pattern leakage is a consequence of thingly uncertainty: as AI technologies infer models from patterns in data, their deployment leads to the projection of such patterns onto the world. Seen through the lens of technological mediation~\cite{verbeek_what_2006,ihde_technology_1990}, events of pattern leakage are a way in which AI technologies make the world legible, often in unintended ways: the things that are `readable' in the world, how we `read' them and what particular `we' is involved here, are shaped through a cascade of ``ontological surprises'' ~\cite{leahu_ontological_2016} which slip into situated realities. It is in this context that we place the metaphor of \textit{entoptic phenomena}.  Benjamin (first author of this paper) became aware of the term through Alastair Reynolds’ science fiction series \textit{Revelation Space}, in which `entoptics' refer to images and animations generated from neural implants~\cite{reynolds_revelation_2003}. The corresponding term entoptic phenomena (Greek for ``within vision'') has real-world origins in the medical and anthropological fields. 

First, entoptic phenomena as used in the medical sense refer to experiential phenomena induced solely from the structural makeup of the eye and date back to the mid-19th century as a concept developed by Helmholtz’s physiological study of the human eye and perception~\cite{helmholtz_handbuch_1867}. They are defined in Oxford’s Concise Medical Dictionary as ``visual sensations caused by changes within the eye itself, rather than by the normal light stimulation process,'' noting that the ``commonest are tiny floating spots (floaters) that most people can see occasionally, especially when gazing at a brightly illuminated background (such as a blue sky)'' (cf.~\cite{noauthor_entoptic_2010}). 
The second, later use of entoptic phenomena is in the anthropological sense as introduced by Lewis-Williams and Dowson. The authors redefine entoptic phenomena as ``visual sensations derived from the structure of the optic system anywhere from the eyeball to the cortex''~\cite{lewis-williams_signs_1988}. Lewis-Williams and Dowson argue that early human cultures deliberately induced altered states of consciousness (e.g. in shamanic rituals) to experience entoptic phenomena, ``scrutinizing [the latter] in the hope of seeing specific forms.'' In this, the authors propose humans developed cultural practices for experiencing such phenomena as a basis for creating geometric or iconic imagery. In the anthropological interpretation, the entoptic metaphor explicitly maps to phenomena such as the wildfire skies mentioned above: human perception is infiltrated by processes that have no direct physical equivalent in the perceived world; yet are influencing how human realities are shaped nonetheless.

\begin{figure}
    \centering
    \includegraphics[width=1\columnwidth]{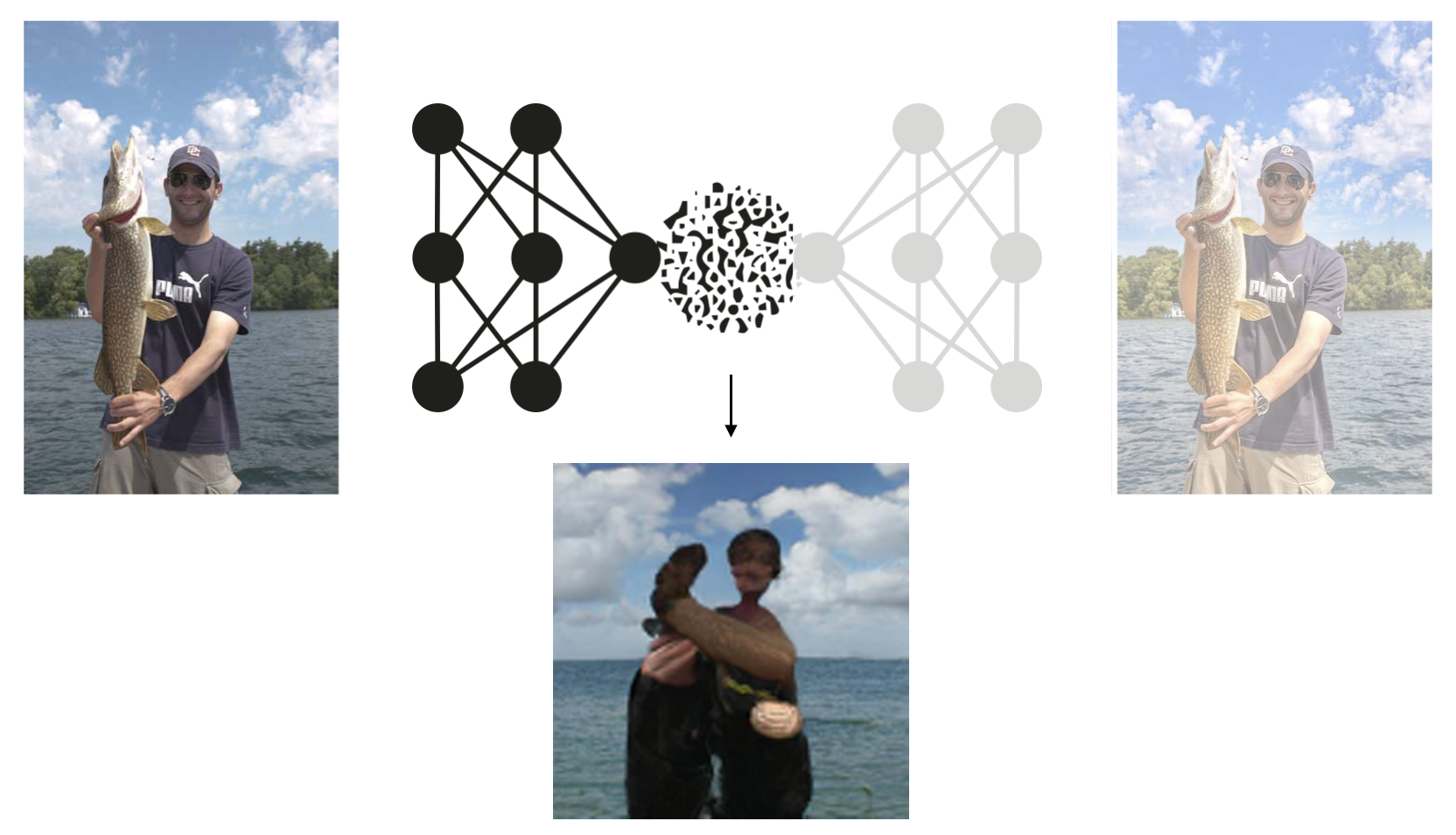}
    \caption{Conceptual diagram for the entoptic metaphor in terms of Generative Adversarial Networks (GANs). In contrast to Figure ~\ref{fig:gan-concept}, learned patterns are not used to improve an input image, but rather constitute the output image themselves.}
    \Description{Conceptual diagram for the entoptic metaphor in terms of Generative Adversarial Networks (GANs); corresponding to Figure ~\ref{fig:gan-concept}. In contrast to Figure ~\ref{fig:gan-concept}, learned patterns (center, abstract representation) are not used to improve an input image (left, same as in Figure ~\ref{fig:gan-concept}), but rather constitute the output image themselves (bottom, synthetic image showing abstract shapes, vaguely mapping to colors and partitions in the input image).}
    \label{fig:entoptic-concept}
\end{figure}

Entoptic phenomena, for our purposes, therefore function as a metaphor for the probabilistic interplay of data-processing (the ``cornea'') and model inference (the ``cortex'') in AI technologies. Similar to the interplay between eye and brain giving rise to entoptic phenomena in visual perception, the interplay between input data and inferred model leads to `entoptic phenomena' in technological mediation. 
Importantly, and in contrast to most prior metaphorical work on AI technologies (see section \ref{sec:metaphors-rtd}), the metaphor is not instructive with regards to a particular purpose. Meaning, it does not resolve to AI-as-X for educational or auditing purposes, but rather to the functional consequences of the material interplay between the components that make up an AI technology---models, algorithms, parameters, datasets. In other words, it is not a metaphor about what AI technologies \textit{are}, but rather what they \textit{do}. In this regard, we frame the entoptic metaphor as a tactic for what Pierce refers to as ``analogical friction''~\cite{pierce_tension_2021}: generating a side-track to current events and capabilities that allows for inquiry. To this end, the entoptic metaphor is conceptualized as a \textit{material analogy} to what an AI technology does below the currently existing threshold of utility; and thereby prompting designerly ways to respond.\footnote{A more extensive theoretical background on the entoptic metaphor can also be found in ~\cite{benjamin_spark_2021}.}

\subsection{Methodological Approach}
This RtD project sought to develop an everyday imaging apparatus powered by AI technologies---the \textit{Entoptic Field Camera}. The Entoptic Field Camera embodies Benjamin and colleagues' concept of pattern leakage (i.e., the shaping of reality through patterns emerging in the material interplay of AI technologies' components), and makes it accessible as a design material. Methodologically, this project is broken into two phases which we present in the following, along with our considerations for analysis. 

\subsubsection{First-Person Design Process}
The implementation of the entoptic metaphor in a concrete artefact was the result of first person research (e.g.~\cite{lucero_sample_2019}) conducted by Benjamin over a period of approximately six months through a process informed by autobiographical design. This approach, i.e. designing for oneself, allows designers to ``rapidly start using and learning about their designs [when developing] exploratory systems that fill a new design niche''~\cite{neustaedter_autobiographical_2012}. Drawing from the entoptic metaphor, Benjamin’s target was a camera application---the Entoptic Field Camera---that embodies what is already happening in AI-driven photography to a more extreme degree. Rather than the rare emergence of how AI technologies shape images of reality (e.g., the wildfire sky example), the application is conceptualized as generating images closer to the source---the \textit{Field}---of imaging. In other words, it deliberately seeks pattern leakage from an underlying `leaky' AI model, overriding the initial photographic representation rather than `improving' it. 

Benjamin wanted a prototype that was mobile and responsive both in terms of interaction as well as output to support photographer’s relations to their environment; allowing for situated reflection and interpretation. Therefore Benjamin decided to make the Entoptic Field Camera run on someone's phone and return AI-generated images quickly. This stands in contrast to the current popular deployment of AI image synthesis models, which generally (i) rely on computational and time intensive solutions to produce outputs and/or (ii) are embedded in systems that are purely \textit{about} using the models directly. As a result, the Entoptic Field Camera was envisioned as a type of research product~\cite{pierce_presentation_2014,odom_research_2016}, manifesting as a refined, functionally and conceptually holistic entity. In this sense, the Entoptic Field Camera is an application that (i) is directly usable and (ii) has an open-ended purpose that is dependent on in situ engagement, inviting users to articulate what taking images with the prototype means. It is therefore conceptualized as a way of discovering the world through AI technologies in a situated, embodied, non-prescriptive and yet purposeful manner. Benjamin’s intuition was that this combination would allow for rich phenomenological interpretations---discerning the world-shaping aspects of AI technologies from how the latter manifest in `worldly' activities (cf.~\cite{heidegger_basic_2005,heidegger_being_2010}).

\begin{figure}[h!]
    \centering
    \includegraphics[width=1\columnwidth]{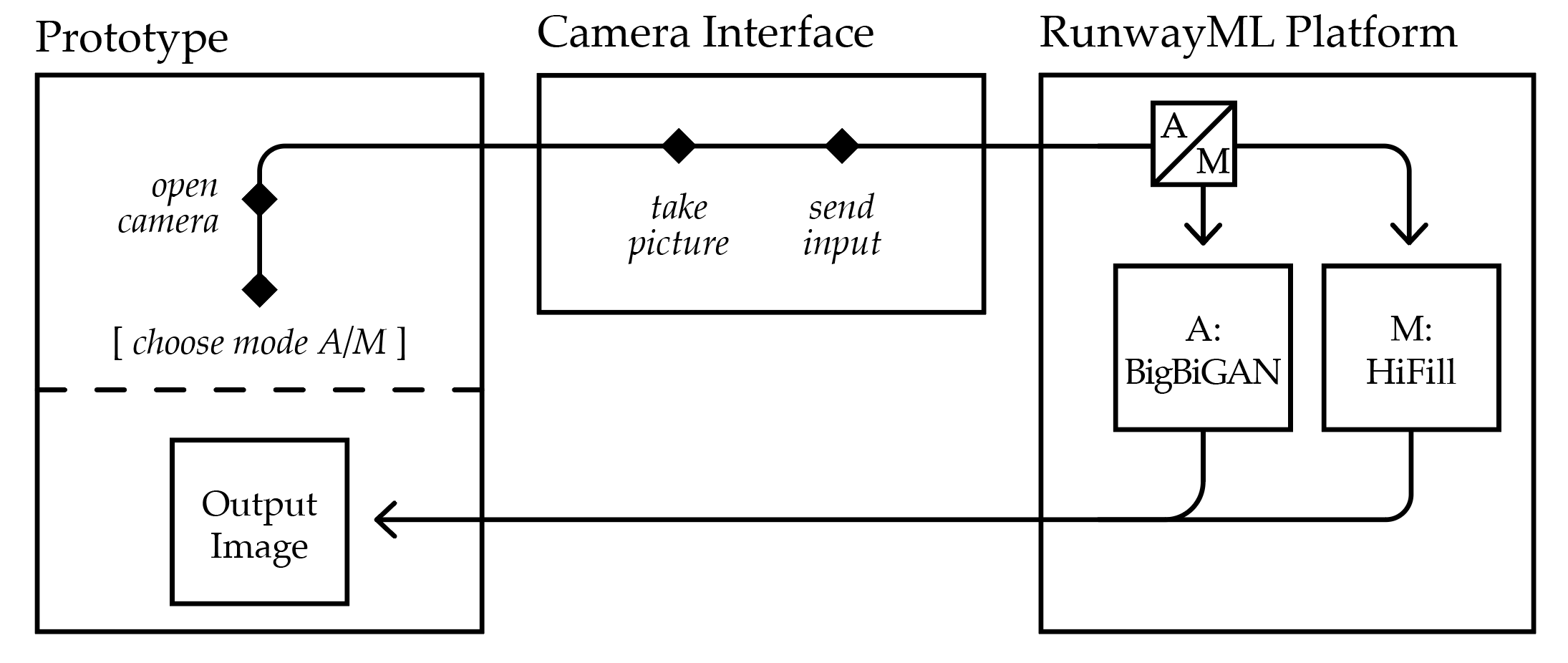}
    \caption{Interaction architecture diagram for the Entoptic Field Camera. A user may choose between Automatic and Manual mode. The first required interaction is opening the user agent camera interface (i.e., the `native' camera of the respective OS). In the latter, a user can take and retake an image. Upon selection, the input image is encoded in base64 and transmitted to the RunwayML platform API corresponding to the selected mode (i.e., either BigBiGAN or HiFill model). The output image is then returned in base64 and rendered in the prototype interface.}
    \Description{A diagram for the flow of interaction in the Entoptic Field Camera prototype, seperating between prototype (i.e., user interface), camera interface, and the RunwayML platform.}
    \label{fig:entoptic-architecture}
\end{figure}

The Entoptic Field Camera is a web application using the online AI hosting platform RunwayML\footnote{\url{https://runwayml.com/}, accessed 11/16/2021.} as the source for output from AI technologies. RunwayML serves as a `hub' for creative projects with AI technologies, allowing users to select among various types of models and host them as application programming interface (API) endpoints.\footnote{Note that the released application, \url{https://entoptic.media/cam/} utilizes a custom-built Jupyter-based backend hosting BigBiGAN and HiFill rather than RunwayML.} The API allows authenticated users to send input (e.g., an image file) to such a hosted model, and receive corresponding output. The application itself is built using JavaScript, with animation events using p5.js.\footnote{\url{https://p5js.org/}, accessed 09/12/2022.} In the version discussed in this paper, the Entoptic Field Camera has two AI technologies that users can toggle between (Figure \ref{fig:entoptic-architecture}), which serve as analogues for `Automatic' and `Manual' modes common in regular camera designs. For the Automatic mode, Benjamin chose the BigBiGAN model~\cite{donahue_adversarial_2016} which has been trained on the omnipresent ImageNet dataset containing ``hundreds of object categories and millions of images''~\cite{russakovsky_imagenet_2015}. BigBiGAN generates an image based on learned patterns in response to an input image. While GANs have recently been somewhat overshadowed by diffusion models, for the purposes of this application it was important that input-to-output processing occurred quickly. Hence, while BigBiGAN's outputs often appear distorted or glitchy, this was seen by Benjamin as an acceptable trade-off given that (i) speed would make the Entoptic Field Camera more product-like, and (ii) the transformation and subsequently necessary interpretation of inputs explicitly aligns with the project goal. Manual mode uses an inpainting GAN called HiFiLL~\cite{yi_contextual_2020}. Inpainting GANs are an application of AI technology that builds on GAN techniques to fill in masked (i.e., hidden from the GAN) sections of an input image by drawing from learned features and the available pixel information. This specific model was trained on the dataset Places~\cite{zhou_places_2018}, comprising millions of images curated for scene (i.e., urban, landscape, etc.) recognition. With these modes, the Entoptic Field Camera incorporates the impact of AI technologies on both the overall appearance as well as content of images. 

\begin{figure}
    \centering
    \includegraphics[width=1\columnwidth]{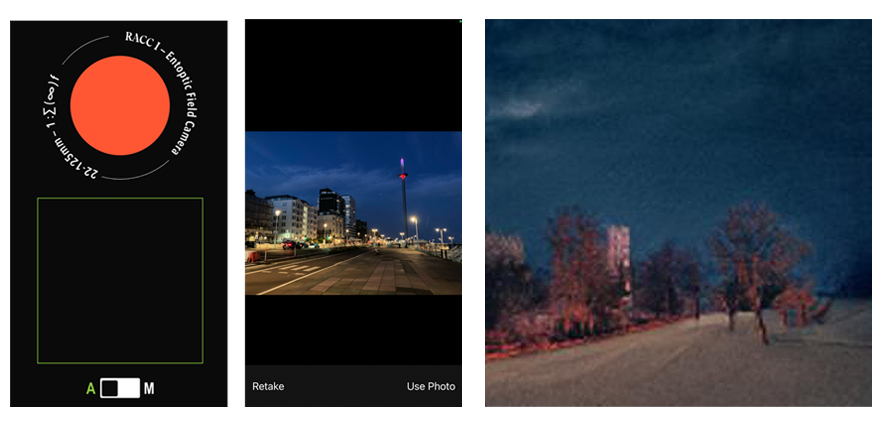}
    \caption{Client-side interaction wiht the Entoptic Field Camera. Left: Interface for the mobile web prototype  as used within the field study. Selecting the red button triggers the user agent camera interface. Middle: Image taken using the user agent camera. Selecting ``Use Photo'' sends this image as an input to the BigBiGAN API since Automatic mode is currently selected. Right: Returned generated image, displayed in the green square on the left.
    }
    \Description{Array of three images that represent the client-side interaction with the Entoptic Field Camera prototype.}
    \label{fig:entoptic-interaction}
\end{figure}

Throughout the development of the application and design of the interface (during the second Covid-19 pandemic winter of 2021/2022), Benjamin first experimented with available models on the RunwayML platform via a browser interface; testing latency and image quality. Once an API was integrated into a first workable interface, these tests where taken out of the home office during daily errands and walks. The design and specification of the application where thus refined in parallel to Benjamin's first findings on the developing entoptic photography practice (see further below). Figure ~\ref{fig:entoptic-interaction} shows a typical interaction with the prototype used during the subsequent field study: upon triggering the user agent camera, a taken image can be passed to the respective mode's GAN (i.e., API for BigBiGAN or HiFill), and a generated image is returned within approximately 10 seconds.

\subsubsection{Field Study with Secondary Users}\label{ssec:fieldstudymeth}
The key motivation for implementing the entoptic metaphor in a concrete artefact was to surface aspects of how AI technologies subtly shape experiences of reality through engagement in a particular practice---here, photography. To this end, the exclusive first-person perspective of conceptualization and making were supplemented: while the conceptual metaphor, design and implementation of the Entoptic Field Camera are predominantly the work of Benjamin, \textit{practices} of photography with the Entoptic Field Camera are then developed and reflected upon in a collaborative manner, with ``mediators [which are] eccentric to the lived experience''~\cite{varela_first-person_1999} of designing the prototype itself. This is an established extension of the autobiographical design method, involving ``secondary users''~\cite{neustaedter_autobiographical_2012}---here, co-authors Biggs, Berger, Rukanskait\.e, Heidt, Merrill, and Pierce---that provide critique and reflection after the fact.

In concert with our position informed by philosophy of technology, i.e. that technologies shape the `legibility' of the world, this method draws on how people develop specific literacies~\cite{bakhtin_speech_1986} through engagement within situated, socio-cultural settings. The importance of such ``real-life literacies''~\cite{purcell-gates_measuring_2012} are key with regards to AI technologies, where \textit{AI literacy} is an ongoing topic of importance in democratic citizenship and education. AI literacy is usually meant as declarative knowledge and competencies of end-users (see e.g. ~\cite{long_what_2020}). In contrast, we are invested in a plurality of literacies (i.e., how the Entoptic Field Camera leads people to `read' the world and subsequently to articulate their readings) with which the impact of AI technologies on experiences of realities, and according design opportunities, can be formulated.

The field study took shape in an iterative and recursive procedure supported by Benjamin and co-authors, who are all affiliated with design research albeit with mixed areas of expertise such as critical making, participatory design, cybersecurity, surveillance, philosophy of technology, speculative design, or design fiction. Through a series of collaborative meetings, the group of authors designed a loose framework for experimentation, which included a minimal technical introduction by Benjamin as well as an introduction to the entoptic metaphor. Each co-author then spent approximately three weeks experimenting and documenting their use with the Entoptic Field Camera in any way they saw fit. Upon completion of the study, authors reconvened to discuss images around a slide-deck and to gather first impressions. Then, co-authors further reflected on their experience in the form of vignette-style reports. The synthesis of findings into the discussion contributions was then led by Benjamin and Biggs (second author).

The field study has received ethical approval by the appropriate body (Ethics committee of Benjamin’s first affiliation); the developed web-application being GDPR-compliant by storing no data on someone's phone, in browser cookies, or the application server. Additionally, the images sent to and received from the respective models are base64-encoded and only stored in the client-side cache, meaning users deliberately have to choose to save an output image. All co-authors that took on the role of secondary users received informed consent regarding these circumstances.

\subsubsection{Analytic Procedure}
Since the phases discussed above differ in method and the roles of the authors (i.e., as creator, participants, analysts), we approach the derivation of findings in a manner that reflects Gaver and colleagues' discussion of emergence in practice-based research. This is fitting, since the project is heavily inspired by an ``anomaly'' (i.e., the California wildfire skies example) and seeks to ``allow technical affordances to suggest new directions'' ~\cite{gaver_emergence_2022}.
Further, through the combination of first-person and secondary perspective methods, this project is also ``a mix of intentionality [i.e., the purposive nature of the Entoptic Field Camera] and openness to change [i.e., not prescribing what findings the former may bring];'' and we therefore paid special heed to manage emergence of possible findings on three levels: Benjamin's individual design process (3.3.1), the shared discussion of the field study among authors (3.3.2), and reflective vignettes written by co-authors as secondary users (3.3.3).
In practice this means that, as mentioned above, Benjamin firstly reflected on their design process from a phenomenological angle that focuses on the emotions, challenges, and perceptions mediated in the design process. To this end, Benjamin kept a daily journal during the process. Secondly, the post-field study discussion was treated as an ad hoc forum to discuss impressions, and was oriented at matters of \textit{form} that the entoptic photography practices took; which was scaffolded by a shared slide deck. Lastly, the textual reflective vignettes of secondary users were considered ``on their own terms''~\cite{gaver_emergence_2022} for thematic analysis carried out by Benjamin and Biggs (second author).

\subsection{Findings}
In this section, we document the findings derived from (i) Benjamin's development and use of the Entoptic Field Camera; (ii) the field study with secondary users (i.e., the co-authors); (iii) a subsequent reflective exercise with all secondary users. To facilitate an understanding of the idiosyncratic and generative approaches pursued in the field study, we show the secondary users' self-selected images. As mentioned above, we frame our findings following Gaver and colleagues’ reflections on ``emergence'' in practice-based research~\cite{gaver_emergence_2022}; and consider the three methodological stages of this project as checkpoints for emergent findings. 

\subsubsection{First-Person Reflections on Design Process}
As the first stage of emergence, the conceptualization, design and development of the Entoptic Field Camera are considered through three themes: the desire for the creation of the Entoptic Field Camera, experimentation with `entoptic selfies,' and the challenge of designing symbolic representations of the entoptic metaphor.

\begin{figure}[h!]
    \centering
    \includegraphics[width=1\columnwidth]{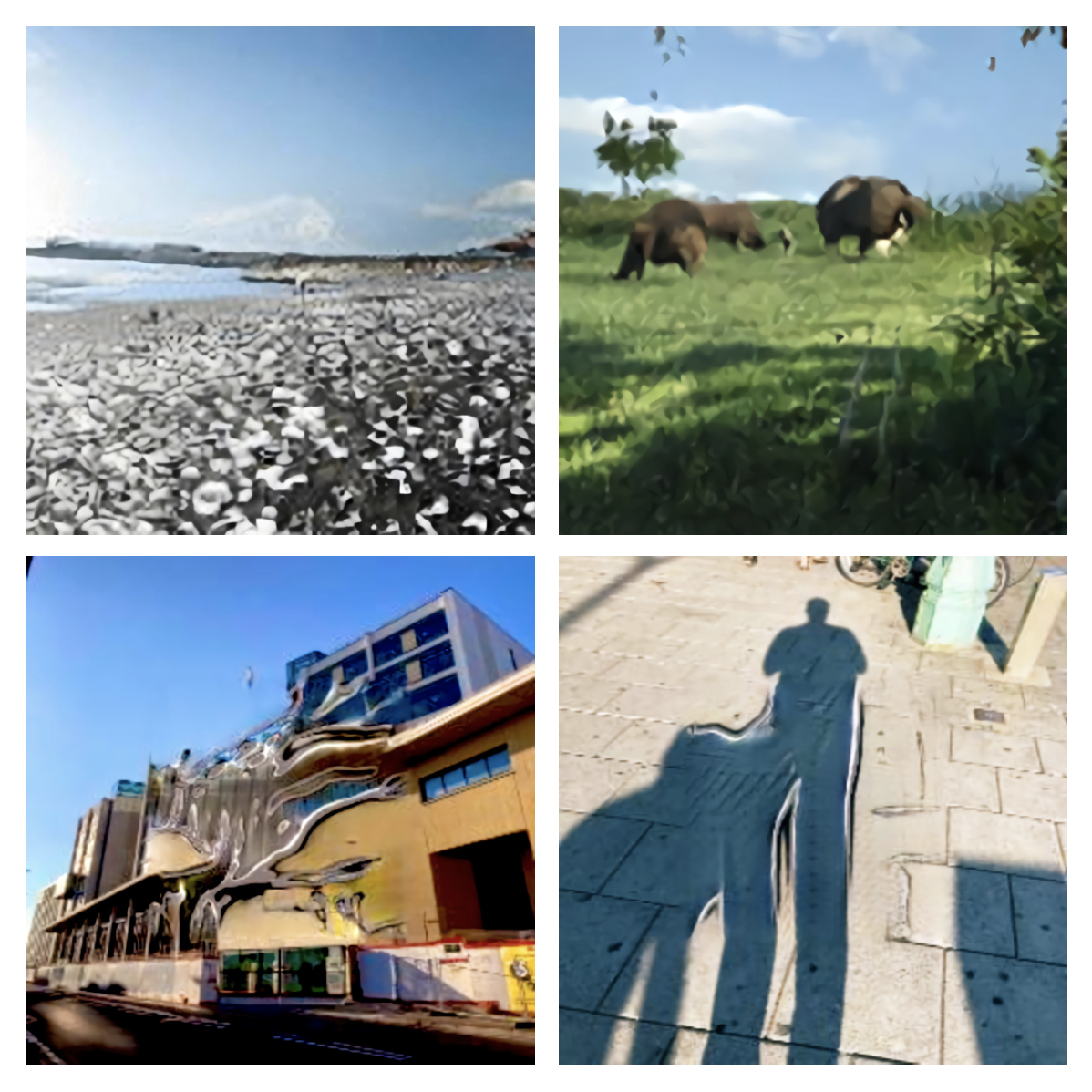}
    \caption{Snapshots by Benjamin during the development phase of the Entoptic Field Camera. The two images on the top depict the Automatic, the two on the bottom the Manual mode, respectively.}
    \Description{4 images made by first author during the development phase of the Entoptic Field Camera. From left to right: 1. An apparent meadow generated by the automatic mode. 2. Unknown animals generated by the automatic mode. 3. Fusion of buildings generated by the manual mode. 4. Fusion of shadows generated by the manual mode.}
    \label{fig:entoptic-desire-exploration}
\end{figure}

\textbf{Desire for and in Making the Entoptic Field Camera --- } During the trial-and-error process of finding suitable and indeed workable AI technologies, Benjamin noticed a particular desire. Two specific experiences in prototyping the two modes of the Entoptic Field Camera are illuminating in this regard (see Figure \ref{fig:entoptic-desire-exploration}). As Benjamin noted, first they found that in Automatic mode, ``I search for the aspects of images that provoke either good attempts at matching the input, or that completely transform what comes in.'' And, when prototyping the Manual mode, Benjamin found that ``I seem to either look to suture (stitch together) things or to erase/camouflage things, either way I’m purposely looking at things and their constellations.'' This desire expressed itself in various ways, though always situated in a temporal frame that put \textit{potential} input and output images into relation. Specifically, as Benjamin learned that the Entoptic Field Camera can alter and reproduce their surroundings in a manner that invites curiosity and reflection, the experience can be described as looking forward to what Benjamin \textit{will have had} to interpret. This persistent searching was reflected in further notes, for example that ``I desire to see what I will have to `catch up to'.''

\begin{figure}[h]
    \centering
    \includegraphics[width=0.9\columnwidth]{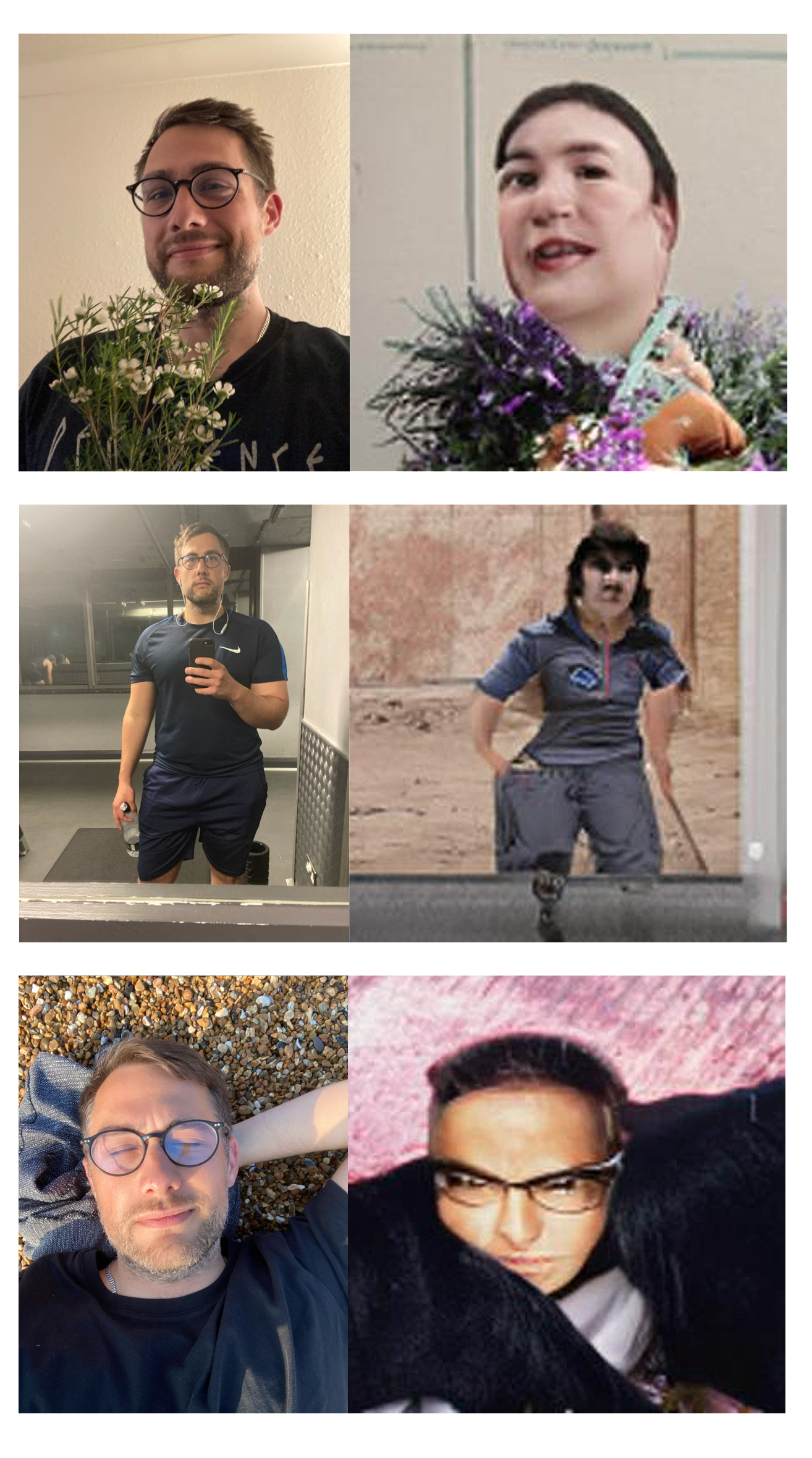}
    \caption{Examples of Benjamin's `entoptic selfies:' the respective input images are  passed through the Automatic mode into synthetic images of selfhood.}
    \Description{3 input-output pairings that represent `entoptic selfies.' On the left of each, the input image is shown: a selfie of Benjamin is various situations. On the right of each, the output image is shown: an entoptic selfie that bear various kinds of resemblance and difference to the original input.}
    \label{fig:entoptic-selfies}
\end{figure}

\textbf{Experimentation with `Entoptic Selfies' --- } A concrete manifestation of the described desire were what Benjamin came to call `entoptic selfies' (Figure ~\ref{fig:entoptic-selfies}). By testing the prototype in everyday situations and practices, the selfie became a way to test how and in which ways feeling, relevance and self-hood persisted in a situated entoptic photography practice. Unsurprisingly, there was no consistency in facial and bodily features between entoptic selfies; each suggesting slightly different identifiers of mood, scenario, fashion or gender. This breadth of self-hood transformations also frequently extended to the background: a gym became an abandoned parking lot, a pebble beach a velvet sky. This became particularly evocative against the backdrop of home office routine. As Benjamin's notes, ``I keep returning to the same place and expect things to be different, yet available for interpretation [such as] my face with closed eyes, lying on the pebbles.'' The recurring theme of going to the same places yet expecting new things had a surprising effect on Benjamin: ``because the [Entoptic Field Camera] is so generative, it leads me to reflect, maybe even worry, how constrained [...] my life is.'' Rather than reflections \textit{on} the AI technologies involved (i.e., BigBiGAN / HiFill), the experimentation with entoptic selfies led to deeply personal reflections.

\begin{figure}
    \centering
    \includegraphics[width=1\columnwidth]{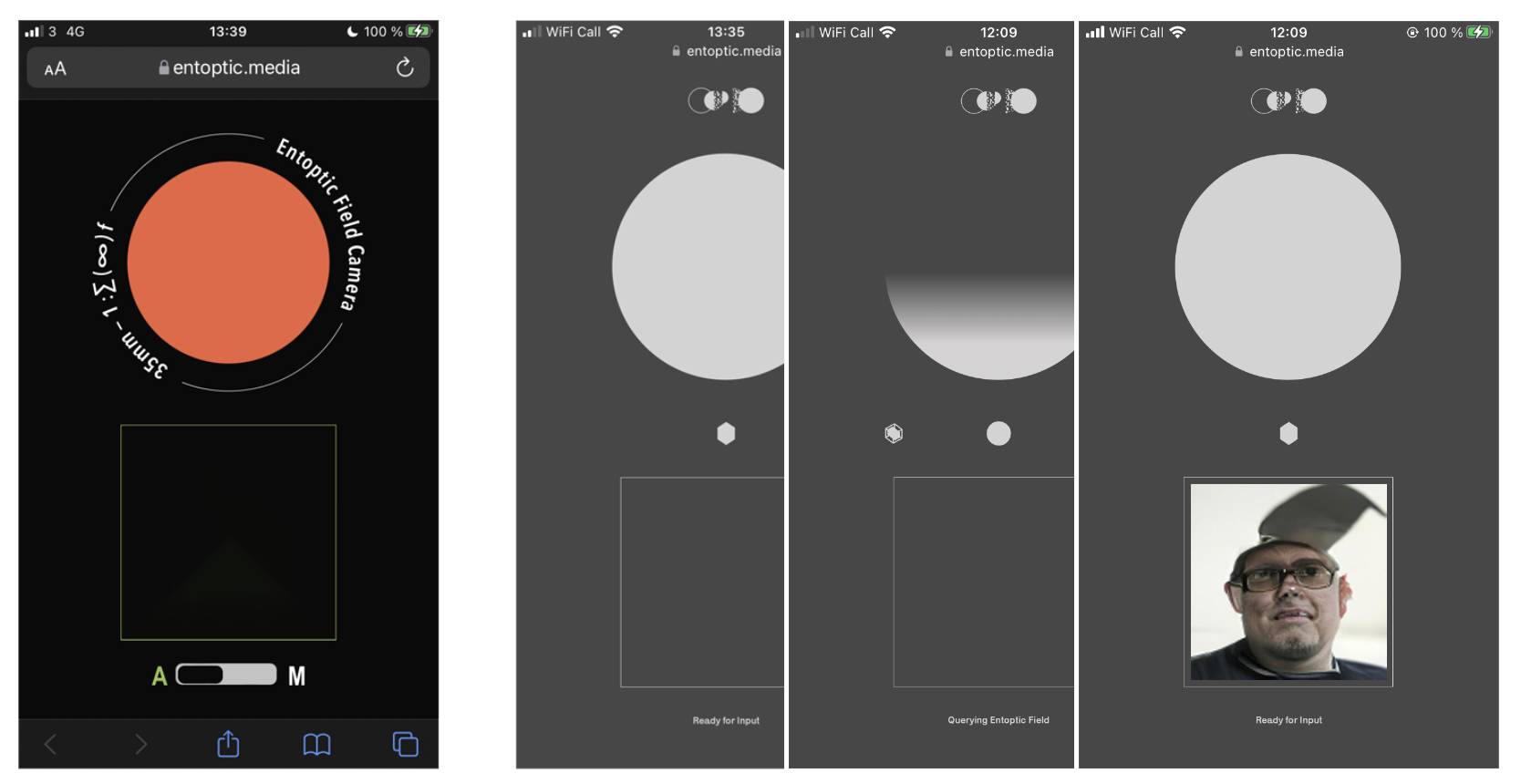}
    \caption{The two interface versions for the Entoptic Field Camera. Left: Prototype as used in the field study. Right: Various states (ready, processing, output) of the later prototype. Note that in this version, the color-space is monochrome rather than recalling a conventional camera aesthetic (i.e., red, green and grey), an abstract representation of the entoptic metaphor has been added (above white button), and geometric animations accompany the data processing step.}
    \Description{Screenshots of the two interface versions for the Entoptic Field Camera. Left: Prototype as used in the field study, featuring a large red button and a `lens-mount' label as well as a toggle switch for the two modes. Right: Various states (ready, processing, outup) of the later prototype. Monochrome colorspace, with geometric animations for loading and an abstract representation of the entoptic metaphor.}
    \label{fig:entoptic-interfaces}
\end{figure}

\textbf{Challenges in Representing the Entoptic Metaphor --- }
In terms of the interface design for the Entoptic Field Camera, Benjamin noted a persistent tension in finding adequate representations of the conceptual import of the entoptic metaphor. On reflection, the symbolic layer formed by using particular graphic elements, typography and semiotic pointers was perceived as potentially obfuscating the basic premise of the entoptic metaphor. To counteract, Benjamin’s first interface design (as used in the field study) purposely mimicked elements of traditional types of photography, and especially referred to the hardware design of cameras (Figure \ref{fig:entoptic-interfaces}, left) such as color-coding, switches, shutter buttons and lens mount decals. Only the latter signified a departure from photography, in that the aperture scale of the Entoptic Field Camera showed as ``1:$\Sigma$($\infty$)f'' rather than, for instance, ``1:2.8f.'' This referred to a GAN’s infinite latent potentials for representing inputs on the one hand, and the fact that each output represented a particular function (i.e., 1:$\Sigma$) on the other. Following the field study, Benjamin’s next iteration abstracted more fully from this design, adding a conceptual graphic element representative of the entoptic metaphor (inspired by Rosenblatt's \textit{Perceptron} illustrations~\cite{rosenblatt_perceptron_1958}), removing the familiar color-scheme, adding an abstract geometric animation for processing input to output, and integrating text clues for the current processing state (Figure \ref{fig:entoptic-interfaces}, right). Nonetheless, Benjamin found themselves consistently in similar binds to work in XAI or FAccT; for instance: how exactly do you develop a visual language for processes that are intrinsically sub-visual? And how do you express a relationship---whether explanatory or suggestive---of visual symbols to such processes? 

\subsubsection{First Impressions of Field Study: Forms of Practice}
At the second level of emergence, an initial post-field study discussion was held online around a slide deck of first impression encounters with the Entoptic Field Camera. Co-authors were invited to upload a selection of images that was representative of their experience, and talk through the practice that they developed during the field study.

In this study discussion, an immediate topic was that two distinct forms of entoptic photography practices had emerged in an uncoordinated way: 1) questions of representation; and 2) a documentary approach. Benjamin, Rukanskait\.e (fourth author), Heidt (fifth author) and Merrill (sixth author) focussed on questions of representation. As outlined above, Benjamin was experimenting with how the representation of the world in entoptic images led to unexpected self-world relations. For Heidt, the study coincided with travel abroad to a cosmopolitan city, a setting they pictured with the Entoptic Field Camera as a ``tool to explore post-human reality,'' a way to look ``at a different world, but never one I would like to live in'' because the Entoptic Field Camera outputs had an ``apocalyptic'' atmosphere. Merrill took pictures of scenarios that they would not ordinarily share, treating the Entoptic Field Camera’s generated images ``as a [cryptographic] hash'' that would obfuscate intimate moments. Nonetheless, to Merrill a feeling of ``intimacy'' still remained, and a ``fear of being interpretable'' took shape concerning the generally nondescript Entoptic Field Camera outputs. Lastly, Rukanskait\.e focussed on individual objects found around the household, ``wanting to recognize'' the original input in the output images. They noted that the Entoptic Field Camera outputs felt like ``pictures from the internet flowing back,'' but in a way that forces one to ``let go of expectations'' due to the unpredictability of what would be flowing back.

The other entoptic photography practice can be grouped as following a documentary approach. Berger (third author), for instance, used the Entoptic Field Camera while on a roadtrip, and reflected on images of this period which on the one hand ``I could not have taken,'' yet on the other still were ``a mimicry of other amateur photographer’s pictures.'' For Biggs, the field study coincided with moving home, leading them to experiment whether ``GANs break a relationship to a space.'' Biggs found that the ``outputs are generic but have a dreaminess,'' and saw themselves confronted with ``pseudo-realities.'' Lastly, Pierce (seventh author) supplemented a long-standing line of research into home surveillance artefacts during the field study, feeding surveillance images (e.g., of the street outside their house) to the Entoptic Field Camera with the goal of ``more images, less curation.'' While to them, the quality of outputs rendered them unusable, they note that ``it seems the [Entoptic Field Camera] tries to picture things that are interesting to an assumed photographer.'' 

\begin{figure}
    \centering
    \includegraphics[width=1\columnwidth]{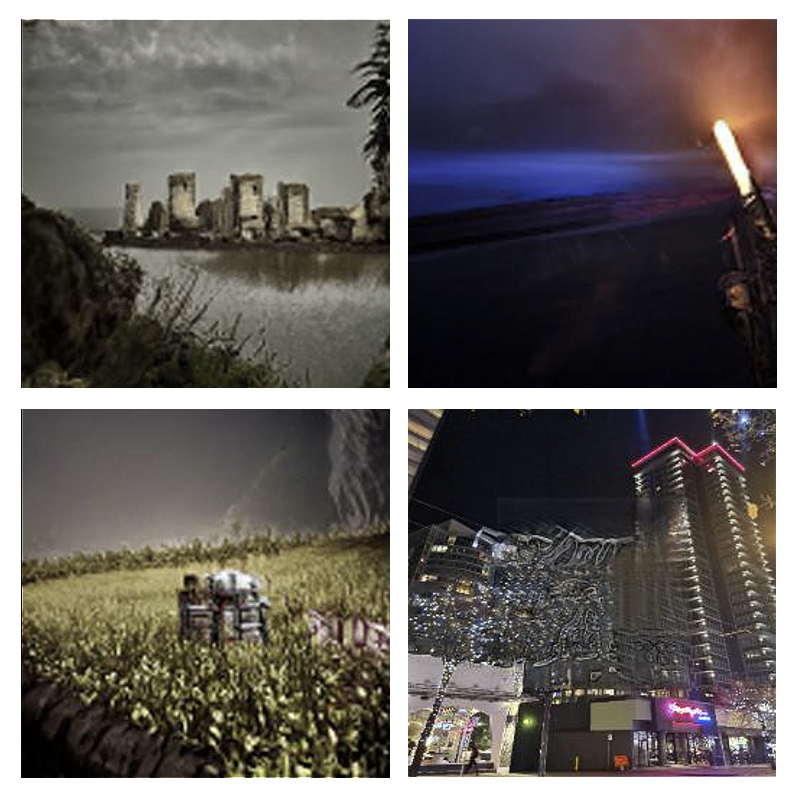}
    \caption{Heidt's presentation of various ``apocalyptic'' and ``post-human'' entoptic images taken while on vacation in North America.}
    \Description{Four images resembling dark decaying cityscapes from the Entoptic Field Camera taken by Heidt.}
    \label{fig:entoptic-5A}
\end{figure}

\subsubsection{Reflective Vignettes on the Field Study: Situated Literacies}
In response to the discussion, a further checkpoint for findings was established that focused less on \textit{what} images co-authors took, and more on \textit{how} those images became meaningful for them. Accordingly, co-authors were invited to formulate their experience in the form of short vignettes. The intervening time, it was thought, would further expose the idiosyncratic utilizations of the Entoptic Field Camera through particular vocabulary choices; which we have referred to above as situated literacies. Without any particular order, we present these vignettes in their entirety in the following due to their richness in ethnographic detail. We also accompany each with notes highlighting significant themes and concepts (in square brackets); since we cannot process all themes as contributions within the length of a paper but are convinced the community could benefit from them. We also accompany each with the self-selected images that the co-authors, as secondary users, chose to present in the slide-deck that facilitated the initial discussion. This presents readers with an overview of the idiosyncratic, generative practices that developed.

\vspace{5mm}

\textbf{Heidt }\textsf{\small \textsc{[ situated technical echo / mirror; interpretation of an alternative world / reality construction; reconfiguration of time; biased by first impressions; perceived `biophobia.' ]}} --- I first tried out the Entoptic Field Camera while traveling Canada for a couple of months in 2021. Specifically, I took the very first picture in Stanley Park, a place quite dear to me, that usually exhibits both a magical and calming effect. However, the entoptic camera did its best to disrupt and pervert said calm. As a test case I chose a nicely framed picture of Vancouver's skyline reflected in the lost lagoon. What the camera threw back at me, however, was more akin to a post-apocalyptic wasteland. Curiously, the image also evoked some dark-romantic associations: skyscrapers appeared as medieval ruins produced by a neutron bomb set off within Vancouver's downtown core. It seemed as if the device had perverted the flow of time in order to fuse a post-apocalyptic future with a desolate and hopeless past. Perhaps, these first impressions induced a certain degree of path-dependency. In any case, afterwards most of the pictures appeared to tell a post-apocalyptic or perhaps post-human story. Interestingly, manual mode behaved especially hostile towards my face, dissolving it into a blob of coldly algorithmic goo. Gradually, I learned that the camera hates all living things, angrily blotting them out whenever they enter the frame. Ultimately, the camera succeeded in constructing its own version of reality, turning every screen it could possess into a twisted mirror which shows life within an alternative post-human universe.

\begin{figure}
    \centering
    \includegraphics[width=1\columnwidth]{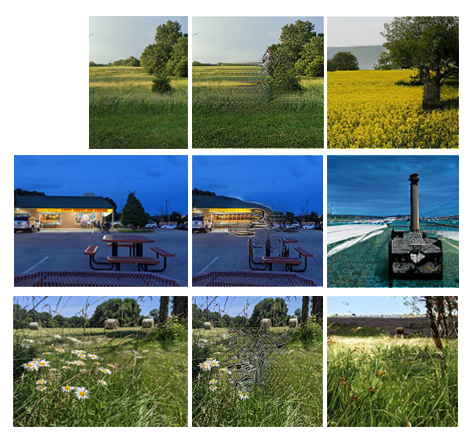}
    \caption{Biggs's presentation of entoptic images taken in the process of moving home. Biggs chose to present the input images as well as respective outputs from both Automatic and Manual modes.}
    \Description{Three groups of three images each. In each grouping, the left-most image shows the original input image by Biggs and to the right, the outputs of both Automatic and Manual modes for that respective input image. The input images are of countryside scenery and lonely diners, with the entoptic outputs partially transforming and resembling the input in a generic manner.}
    \label{fig:entoptic-2A}
\end{figure}

\vspace{5mm}

\textbf{Biggs }\textsf{\small \textsc{[ multimodal, dreamlike quality of images; materiality of GANs; subjective connection to images; question of authentic relations to the world; difference between recollection and record ]}} --- I used the blurry, other-worldly indeterminacy of the networked images it creates to my advantage to reflect on the perception of place that can shift when one listens deeply or when one reflects in memory. I started exploring the Entoptic Field Camera around the time I was about to move. Therefore, I hoped that the images would take on the dreamy quality of memory and the quality of the images that appear (blurry/atmospheric) when one listens to sounds of a place. 
Using the Entoptic Field Camera with images tied to specific memories, a specific place, and a certain nostalgia made me wonder if the Entoptic Field Camera generated images were any longer ‘proof’ of my being there – if they any longer corresponded to place – and if so to what degree. 
On the one hand, the technical and aesthetic meaning built through using the Entoptic Field Camera inspired reflection and creative exploration around the themes of other worlds, memory, and the senses, but on the other hand, the Entoptic Field Camera unsettles expectations of the camera to be a recording device, proof of experience, proof of travel, garnered through the mechanical objectivity of the camera (Susan Sontag, on photography). I walked around my neighborhood I lived and captured a few simple images and collected the soundscapes they correspond to. First was a photo of a farm field with baled hay the sound of cicadas droning in the treetops, second was a field being buffeted by strong winds and a tornado alarm, and third, an ice cream shop in a strip mall at twilight and the sounds of people sitting outside laughing and chatting and cars going by. I strung together the images and sound into a video where the sounds are quite crisp and clear but the images blend into one another. I thought perhaps the crisp, situated sound, paired with the shifting and indeterminate visual field reflected a kind of tension between recollection and record.

\vspace{5mm}

\textbf{Rukanskait\.e }\textsf{\small \textsc{[ mundane objects and their representation; recognition, transformation and destabilization; returns from the uncanny; images made of images; intrusion; lack of control regarding foreground / background; decentering of intention ]}} --- I took photos of objects at home I thought the camera might recognize or be able to transform. Taking pictures with the camera meant having everyday objects reflected back in an unstable, uncertain manner. Or not having them reflected at all, and instead taken for something else; something weird, uncanny. An abstracted mash of anonymous internet images. Their alienness became obvious when I noticed what looked like a date stamp in the lower right corner of one reflection. I could not control what the camera picked up on – from its point of view, the objects, colors, textures, and shapes I considered to be background became central. I kept forgetting the camera picked up on everything in the image frame, large aspects of which I did not consciously focus on. At the same time, it felt encouraging to see the things I photographed reflected back, even if distorted; even if the same thing never reflected in the same way once, I searched for and sometimes found repetition in their reflections. Towards the end of exploring the camera I stuck to photographing abstract patterns, for example, I took close-up photos of plant leafs: no intention or expectations, just pure curiosity. The red corona-virus-looking thing on what looks like a sink remains a mystery. Could it be a result of the noise in the model’s training data? Reminds me of the absurd things that showed up in other field studies, for example, the donkey-like animal prompted by a park fountain.

\begin{figure}
    \centering
    \includegraphics[width=1\columnwidth]{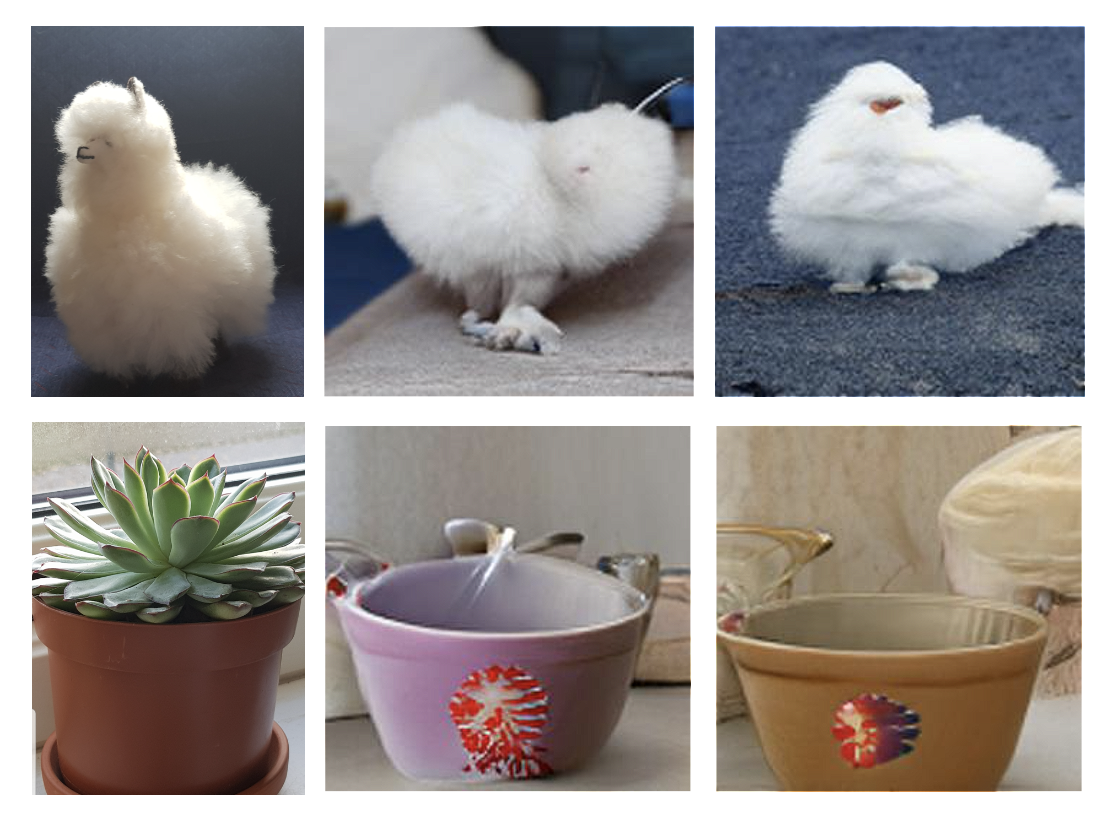}
    \caption{Presentation by Rukanskait\.e of various experiments on representation of mundane things in entoptic images.}
    \Description{Three groupings of images. On the top left, a woolly toy lama is shown as the input image for the Entoptic Field Camera, and two Automatic mode outputs shown besides that roughly resemble the appearance of the lama. On the bottom left, a succulent in a plant pot is shown with two Automatic mode outputs besides, which maintain the appearance of a plantpot but do not show the succulent.}
    \label{fig:entoptic-4A}
\end{figure}

\vspace{5mm}

\textbf{Berger }\textsf{\small \textsc{[ reality shifting as a `dreamlike adventure'; distributed agency; infinite mirror; hopeful misinterpretations; hallucinations / dreamlike aspect; future reader ]}} --- 
Sitting through yet another pandemic induced video call, I hoped and longed for the EFC to computationally lift the pandemic travel ban and take me on a dreamlike adventure. Much of my everyday reality is computationally augmented, I am listening to noise reduced 3D-audio through my headphones and my car politely countersteers my driving. Now, how would a computational road-trip hold up? With road-trips one can watch the world go by from a safe distance. In driving endlessly, I composed street, field, and sky over over and over again and EFC extended this banality beyond the horizon. We created about 100 images and I can’t quite tell who the photographer is. It was as if we never left the realm of what the algorithm believes 100\% of road-trips look like. Likewise, the GAN model did also serve as an infinite mirror for roads I could have taken, or may have had. The computational road-trip shields the viewer from any unpleasant surprise. We also ended up in the woods, where most plants were plagued, damaged, corroded. This is where, counterintuitively, the EFC created hope. Through the camera I asked, how AI sees and interpret common tree diseases. The GAN seems to have computationally competed against decay and both confidently and consistently reversed any ill effects. The camera made me hallucinate two worlds that could be. Will nature in a distant future be only computationally healed? And can you, future reader, see the road I took?

\begin{figure}
    \centering
    \includegraphics[width=1\columnwidth]{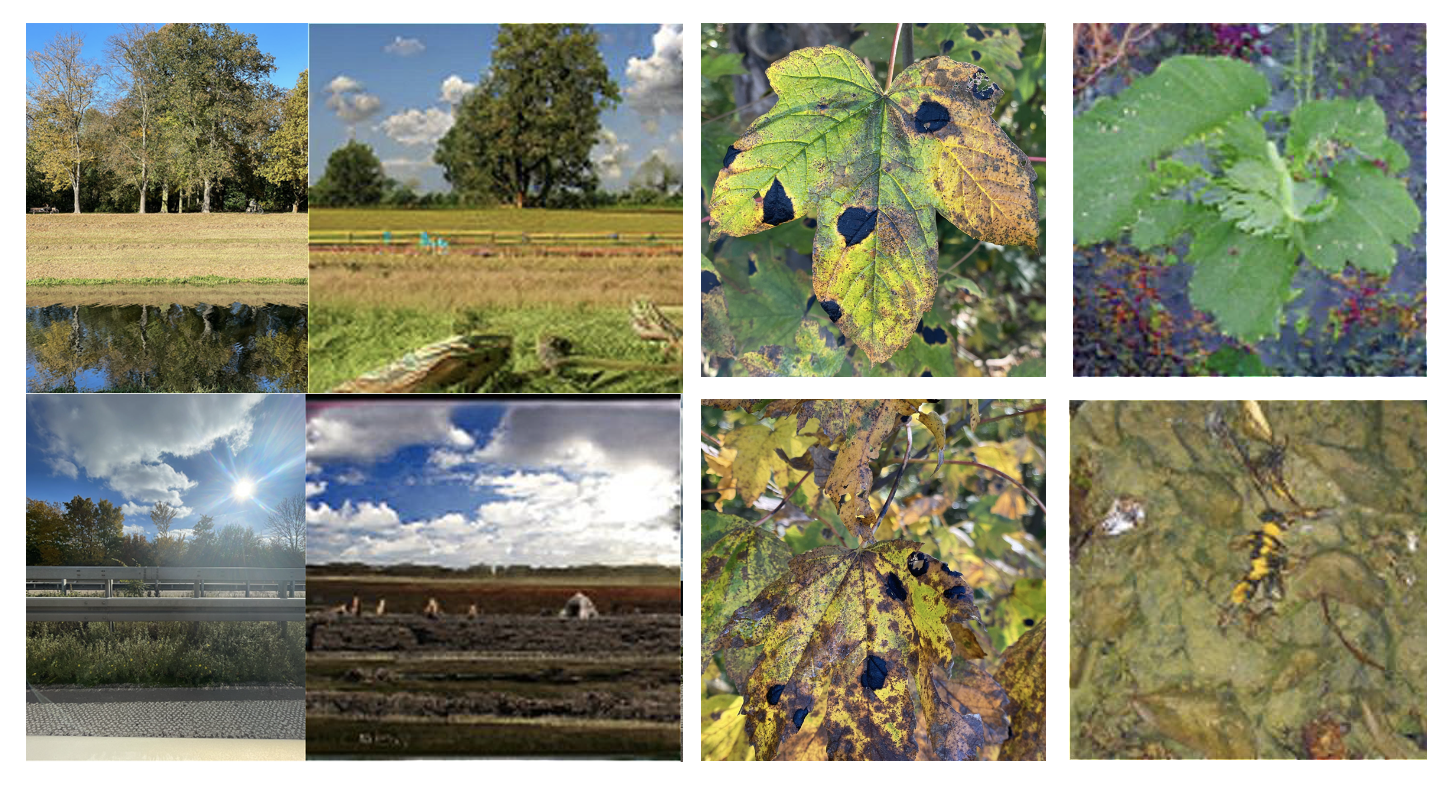}
    \caption{Selection from Berger's ``computational roadtrip'' (left), and forest excursion (right).}
    \Description{Two groups of images, showing input and Automatic mode output images. On the left, input images show roadsides as if taken from a car window and their respective output images that transmute these into vaguely matching landscape scenery. On the right, input images show decaying leaves that appear `whole' in their respective output images.}
    \label{fig:entoptic-3A}
\end{figure}

\vspace{5mm}

\textbf{Merrill }\textsf{\small \textsc{[ intimacy and obfuscation; anonymity through distortion; anonymity permeated by intimacy; questionable permanence of privacy; future reader ]}} --- 
I took photos inside my home, photos of intimate moments, moments I wouldn’t typically share. I expected the camera to obfuscate these moments: to make them less personal, easier to share. In a sense, it did. It anonymized them, making specific places, people and situations difficult to discern. In another sense, it didn’t obfuscate the fact that these moments were intimate. All of these photos evoke to me - perhaps because I know their story - a privateness. Even as they appear now, I admit I feel as if I might be overdisclosing by sharing them. Today, I wonder if some future technology will be able to ``reverse'' these photos. I thought of the camera as a sort of hash function, a trapdoor through which the source image could not be recovered. Only once I took and uploaded these photos did I realize: I have no guarantee that the camera is a trapdoor. There is no mathematical assurance (again, gesturing to hash functions) that these photos are irreversible. Will you, reader of the distant future, be able to recover the original images from these? Will you choose to? If so, why? What would it mean for you to do so?

\begin{figure}
    \centering
    \includegraphics[width=1\columnwidth]{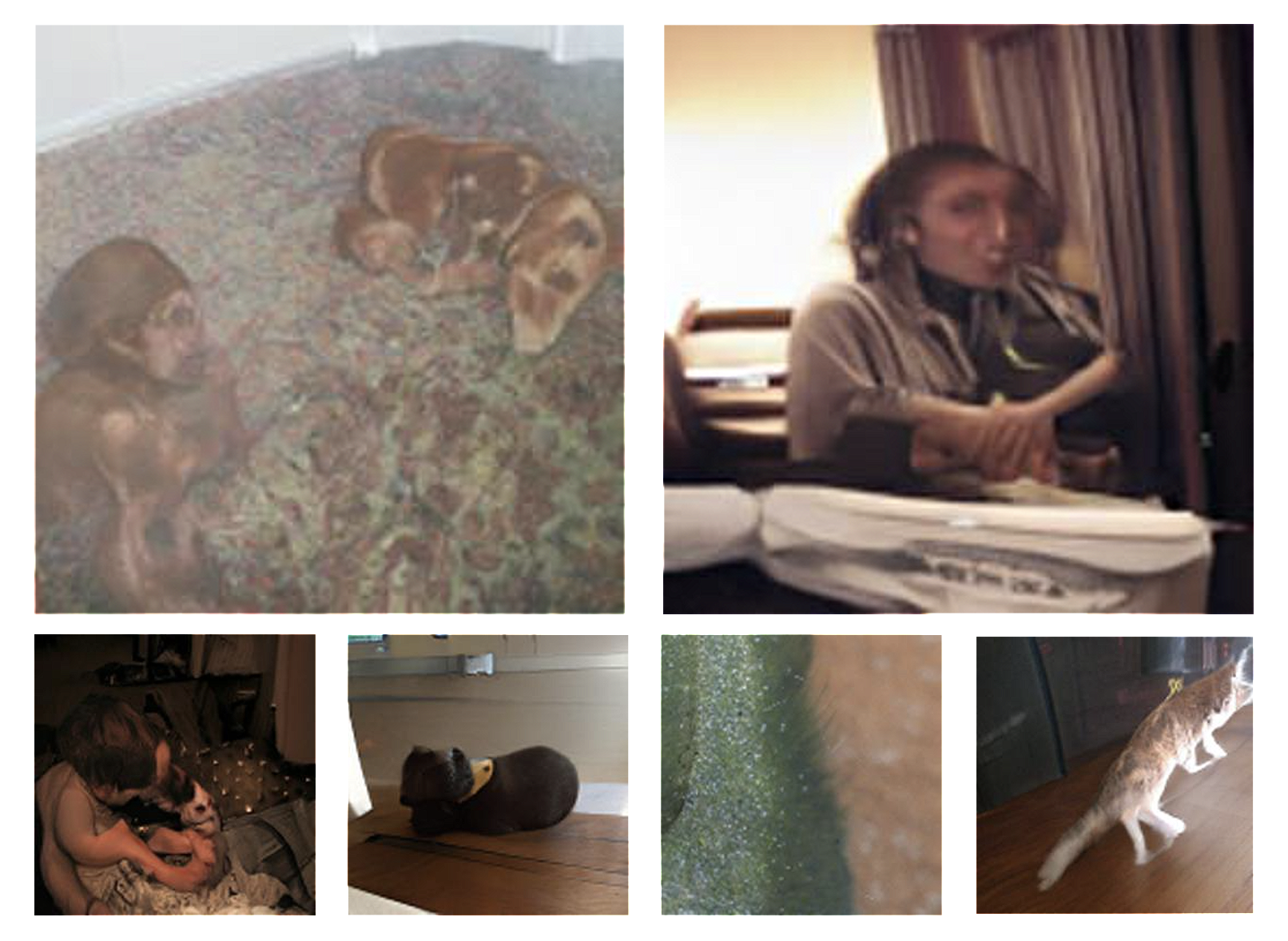}
    \caption{Presentation of Merrill's `intimate' entoptic images.}
    \Description{Six images showing Automatic mode outputs exclusively. All have an uncanny appearance, featuring beings and one definite human in a home setting.}
    \label{fig:entoptic-6A}
\end{figure}

\vspace{5mm}

\textbf{Pierce }\textsf{\small \textsc{[ distribution of agency; image selection; Rrcursive chains of image processing; meta-imaging; distancing from subjects; intrusion; lack of control regarding foreground / background ]}} --- 
I set up a Nest smart security camera affixed with a zoom lens and aimed the camera at an area of approximately 1 square meter in the alleyway outside my home. I then reviewed the ``motion'', ``person'' and ``animal'' detection events curated by Nest Aware. I selected several images and fed them into the entoptic camera. I then recursively fed some of the image outputs back into the entoptic camera. Results were mixed. More often than not, the camera seemed confused and unable to correctly interpret the image. Not surprising, perhaps, given that the images were unconventionally framed and only captured fragments of the subject. I was curious if the entoptic camera might be able to infer the entire subject from a component. In most cases it did not. I also was curious what happens with recursive images. Second and third order images seem to introduce additional entropy, and stray further from ``the subject.'' Viewed relative to the background or texture though, they perhaps fared much better. Conceptually, I also was intrigued by the chains of dependency and mixture of human and machine agency: a camera pointed at a small random area of a thoroughfare, with a smart camera system that captures and selects a subset of events, then I curate a subset of these, then I feed them into the entoptic camera, then I feed the outputs back in, ... My personal favorite was the motley complete child’s face that morphed into a partial face of Kermit the Frog. 

\begin{figure}
    \centering
    \includegraphics[width=1\columnwidth]{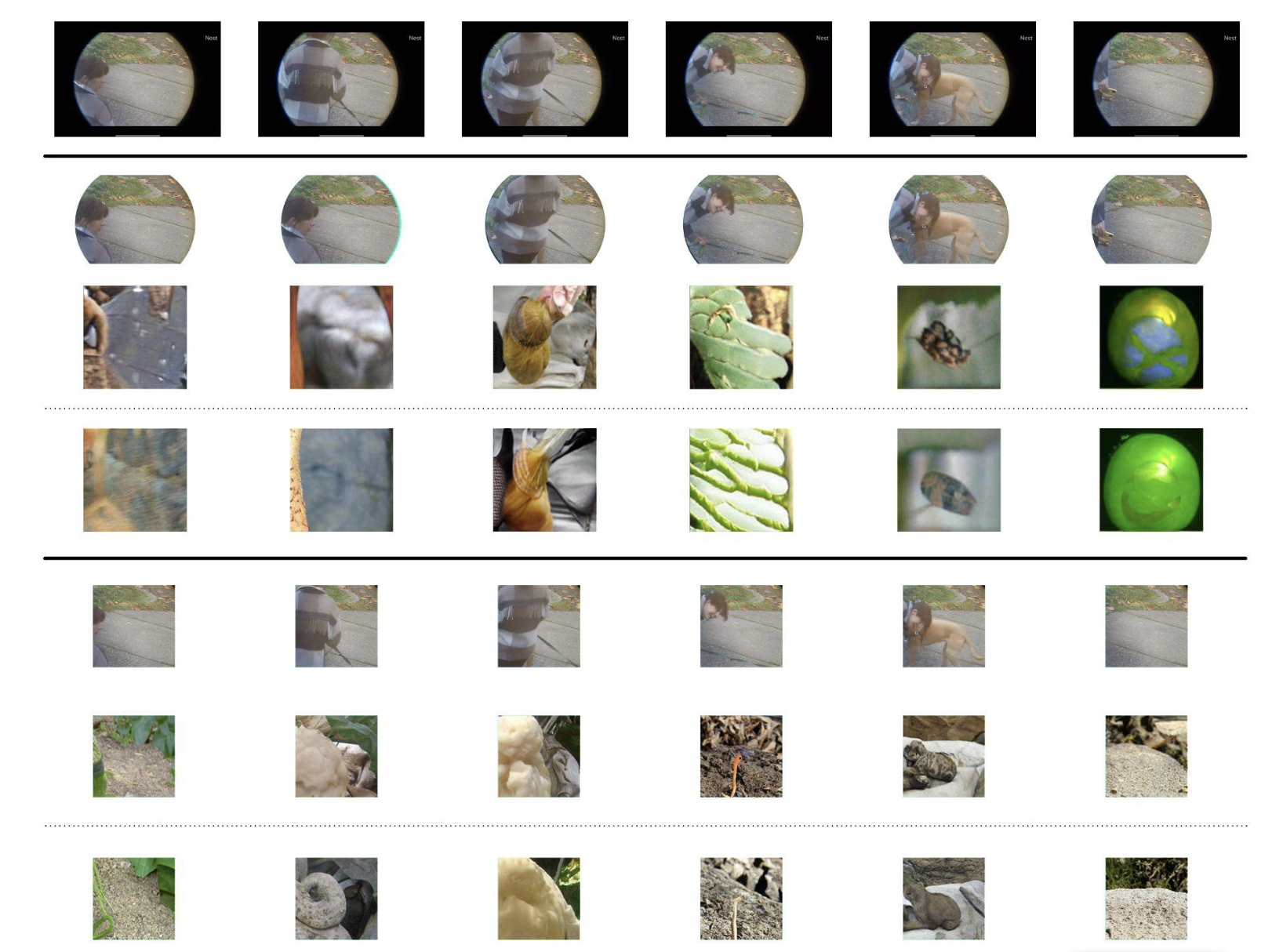}
    \caption{One of Pierce's presentations of multiple input-output pairings; i.e., initial Nest smart security images that are fed into the Entoptic Field Camera over various iterations.}
    \Description{A 6 by 7 matrix of input and output images. Each column is one row of recursive imaging using both the Nest smart security system as well as the Entoptic Field Camera, with images becoming progressively more abstract.}
    \label{fig:entoptic-7A}
\end{figure}

\section{Discussion}
In the following, we outline three major contributions of this project. First, drawing from the development, design and usage of the Entoptic Field Camera, we outline design implications that further enable established design methodologies to engage AI technologies. Second, we outline the conceptualisation of an RtD research space in relation to AI literacy. Lastly, reflecting on the relationship between materiality, design practice and affordances, we outline a research trajectory for a more comprehensive inquiry into the relationship of the affordances of 21st century design and AI technologies.

\subsection{Design Implications of the Entoptic Field Camera}
We argue that there are numerous implications for RtD methodologies and programs where the entoptic metaphor, whether concretized in the Entoptic Field Camera or other `entoptic media,' may be deployed. Below, we gather these around themes which reflect methodological propositions and practice-oriented provocations. In sum, these implications highlight that the entoptic metaphor allows design research to shift from questions \textit{about} `AI' to how the material arrangement of actual AI technologies shapes experience and foregrounds particular concerns and opportunities for design.

\subsubsection{Materializing Distributions of Agency}
The question of causality, dependencies and mixed agency is a vital one with regards to AI technologies, which prove stubborn in various domains such as the socio-economic conditions of crowdworkers~\cite{irani_turkopticon_2013}, extractive end-user engagements~\cite{ekbia_heteromation_2014} or intellectual ownership (see e.g. ~\cite{robertson_us_2022}). The entoptic metaphor, and the findings from its concretization in the Entoptic Field Camera, can be used to engage this complex through critical ~\cite{pierce_expanding_2015,bardzell_what_2013} and reflective~\cite{sengers_reflective_2005} design which seeks to unfold or question how agency is distributed with regards to AI technologies and the systems these are embedded in. Merrill was most explicit here, foregrounding their inquiry into ``chains of dependency and mixture of human and machine agency'' in which the Entoptic Field Camera allowed them to probe the boundary conditions of such distributions. The notion of testing boundary conditions is of critical importance in work on AI technologies that foregrounds ethical and power-related issues stemming from the relationship of models and datasets (see e.g. ~\cite{birhane_multimodal_2021,uday_prabhu_large_2020}). We argue that RtD projects such as the one presented here could open up the usually expert-constrained methods of such critical auditing. 

As an example, a future iteration of the Entoptic Field Camera or similar entoptic media could be used for ``grassroots documentaries''~\cite{green_enabling_2017} in which members of specific communities document their everyday or a specific event through such an artefact. As we noted above, both GAN techniques (i.e., BigBiGAN and HiFill) were trained on specific datasets--the ubiquitous ImageNet and Places, respectively. Biggs and Berger in particular noted the ``generic'' nature of images that foregrounds a dream-like (dis-)connection as well as strangely reassuring aspect at the same time. Accordingly, entoptic documentaries could, for instance, focus on how a community distinguishes itself from or recognizes itself in the abstract-generic nature of entoptic imagery (e.g., in post-hoc discussions); or reflect on questions relating to privacy and intimacy which may be sensitive or close to heart for particular communities. The entoptic metaphor, we propose, can thus be employed to study  the boundaries and fault lines in the relations between an AI technologies' material interplay and the concerns of particular communities.

\subsubsection{Dematerializing Privacy Concerns}
Merrill's reflection on the relationship between anonymization, obfuscation and intimacy opens up a further trajectory for critical design. Merrill expressed that they ``thought of the camera as a sort of hash function'' which would allow them to share subjectively felt intimate moments. This suggests a layer to entoptic photography which may be oriented around questions of privacy. While it is doubtful that an Entoptic Field Camera would be useful for any conventional practice of targeted surveillance---as Pierce found by using actual Nest security images as input---precisely this can be interpreted as a utility of a different kind. Bellanova and Fuster have described how airport security body scanners needed to be outfitted with ``dematerialization operations''~\cite{bellanova_politics_2013} following complaints over violations of privacy; leading to a downgrading of their technical capacities to only reveal abstract, generic bodyshapes on scanning results. An entoptic surveillance application could arguably perform a similar service, `dematerializing' the initial input image while generating a still causally and materially related output image. A potential prototype could be designed as an `entoptic monitor' of a specific location, and show only output images---which, given the GAN's radical transformations, may prompt questions that echo Merrill's reflections on the nature of privacy, intimacy and obfuscation; and Pierce's experiments with input-output recursivity: what `remains' of the emotional connection or political import of being surveilled, when surveillance images do not resemble their input's \textit{appearance} yet are undoubtedly \textit{materially} connected? In this regard, the entoptic metaphor can support adversarial design (cf. ~\cite{disalvo_adversarial_2012}) which makes the presence of AI-driven surveillance both readily apparent (see e.g.~\cite{noauthor_shuttercam_2022}) as well as provoking reflections on its transformative effects, to seed discussions and prompt reflection through (de-)material sensitization.

\subsubsection{Experimenting with Forms of Subjectivity}
Conceptualizing the Entoptic Field Camera's---or rather, the GANs'---dematerialization of what is given as input can be taken up further through reflective design~\cite{sengers_reflective_2005}. Sengers and colleagues introduce the latter as a methodology that foregrounds ``reflection on unconscious values embedded in computing.'' Aspects of this stance can be discerned in an literal sense throughout co-authors’ experiences. Rukanskait\.e for instance noted that ``the things I photographed reflected back in an unstable, uncertain manner;'' and Berger and Heidt both referenced mirroring. We argue that such notions open a critical avenue to reflect on potentially constraining as well as emancipatory potentials embedded (through data and algorithmic bias) in AI technologies. With regards to the latter, an example can be found in Turtle’s critical self-portrait engagement regarding ``queer becomings with AI''~\cite{turtle_mutant_2022}. This potential for GANs, and generative AI technologies more generally, to allow for experimentation on opportunities as well as boundaries for self-expression was found explicitly in Benjamin’s `entoptic selfies.' This brings up immediate opportunities for future critical-reflective work: what exactly is the distinction between entoptic selfies and more common forms of online self-presentation? What is missing on either side of these forms---do entoptic selfies lack verisimilitude, or do they express something about the technologically mediated contemporary lifeworld that Instagram stories lack in turn? And: just \textit{who} is the photographer in either case?\footnote{That the Entoptic Field Camera is generative of exactly this kind of reflection received some informal confirmation during its exhibition at the 2022 Ethical Dilemma Cafe as part of Mozilla’s Mozfest; see e.g. \url{https://cubicgarden.com/2022/05/08/mozilla-bbc-ethical-dilemma-cafe-manchester/}, accessed 09/06/2022.} 
\subsubsection{Decentering Conventional Scales of Representation} More-than-human design and sustainable HCI (SHCI) are growing areas of research, frequently confronting the implicit anthropocentric narcissism of emerging technologies (see e.g., ~\cite{reddy_encountering_2020,nicenboim_more-than-human_2020,wakkary_things_2021,jonsson_doing_2022,homewood_removal_2020}). A key strategy in efforts to transcend the extractive logic of established user-/human-centered design is the \textit{decentering} of a human-exclusive logic means and ends ~\cite{wakkary_things_2021}. Regarding the latter, our framing and findings suggest that the entoptic metaphor as applied to photography may suspend this logic to more fully reflect how AI technologies affect experiences of reality. Specifically, Rukanskait\.e noted how they ``kept forgetting the camera picked up on everything in the image frame, large aspects of which I did not consciously focus on.'' Pierce's observation of the ``additional entropy'' through recursive input-output pairings is also related; suggesting practices of decentering can be entoptically augmented to probe for more-than-human notions of representation. An example can be found in Berger's forest excursion, where they noted that ``the GAN seems to have computationally competed against decay and both confidently and consistently reversed any ill effects,'' and further asked whether ``nature in a distant future be only computationally healed.'' The superficial removal of decay echoes work that centers on removal or undoing (see e.g ~\cite{homewood_removal_2020,jonsson_doing_2022}), and introduces a further twist: through entoptic imagery, what appears as \textit{signifiers} of anthropocentric effects (e.g., decay, erosion, etc) may be removed, or rather, made mute through the generativity of AI technologies. Inversely, Heidt's perception of ``apocalyptic wastelands'' are shaped by the \textit{introduction} of signifiers. 

This transformation of what is represented in photography does also not need to be restricted to signifiers in space, but may also relate to \textit{time}: Heidt argued that the Entoptic Field Camera seemed to ``pervert the flow of time'' by converting a present scene into its correlated future wasteland. Such perceptions recall Biggs and Desjardins' \textit{High Water Pants}~\cite{biggs_high_2020} project, the titular artefact being computationally-enhanced pants which react to their wearer entering a future zone of flooding by shortening the pants leg. Biggs and Desjardins see their prototype as an ``oracle [which] bends time'' by enabling situated interaction with possible futures in specific practices (in their case, cycling). Similar to the effect on notions of selfhood we described above, the Entoptic Field Camera can be interpreted as a time-bending artefact: de-materializing present foci, and re-materializing potential futures. Either way, the important point here is that the removal or addition of signifiers do \textit{not} occur either at the behest of subjective choice of motive nor technical `intelligence,' but rather through the interplay of AI technologies' models and datasets in response to a circumstantial arrangement of pixels. Entoptic photograpy may thereby undermine prevalent anthropocentric notions of what make suitable objects of photography; and disclose potentials for decentering design practices accordingly. 

\subsubsection{Exploring Ludo-Entoptic Play} 
Lastly, as shown in our findings, both Biggs and Berger picked up on a dreamlike or hallucinatory potential of the Entoptic Field Camera, while Rukanskait\.e referenced the uncanny quality of the images. This recalls the anthropological interpretation of entoptic phenomena offered by Lewis-Williams and Dowson~\cite{lewis-williams_signs_1988}: the deliberate search for entoptic patterns that are then concretized in forms of cultural expression (e.g., cave paintings). In terms of AI technologies, the entoptic metaphor accentuates exactly this generativity in terms of an inter\textit{play} among components (e.g., model, data, algorithm). This suggests some productive overlaps between ludic design~\cite{gaver_drift_2004,gaver_indoor_2013} and the Entoptic Field Camera. In this overlap, ludo-entoptic artifacts could be designed that thematize the ``intoxication of creative play''~\cite{flusser_into_2011} which AI technologies such as image synthesis models are already bringing about. 

Aligning situated play with withdrawn interplay, ludo-entoptic artefacts expand the human-artifact locus (e.g., person + device coupling) towards entanglements with the components of AI technologies. Examples could be Entoptic Glasses or Contact Lenses, conjuring up possible entoptic hide-and-seek or games of tag; where pattern leakage intervenes in ordinary searching, hiding and anticipating practices. The ludic aspect of entoptic media, then, is of a different kind than the creative play currently seen with prompt-based image synthesis models (e.g., Dall-E, Stable Diffusion, etc.). Whereas AI artists Herndon and Dryhurst advocate for such AI technologies ``[as] a tool for jamming, rapid iteration and potentially co-authored social experiences''~\cite{herndon_infinite_2022}; the Entoptic Field Camera’s creations do not directly correspond to supposed 1:1 relationships in human-AI collaboration. Rather, ludo-entoptic artefacts embed playful creation within an ``infinite mirror'' (Berger) that disregards intended focus or distinctions between background and foreground ([4A, 6A]). Both the Manual as well as Automatic modes, i.e. synthesis and inpainting GAN techniques, used in the Entoptic Field Camera can serve as impulses here: whereas an inpainting ludo-entoptic artefact may allow, force or train people to use peripheral vision for play; a synthesis version could play on contemporary imaginations of multiverses or augmented reality applications.
\subsection{RtD between AI Materiality and Literacy}
As we have argued above (see Section ~\ref{ssec:fieldstudymeth}), we interpret AI literacy on a more fundamental level than declarative knowledge and competencies of end-users (see e.g.~\cite{long_what_2020} for an overview); not only referring to technical knowledge, exact terminologies or end-user competencies, but rather as also referring to the intuitive (mis-)understandings and experiences of how AI technologies shape our experience of the world. Accordingly, there is a potential gap between what is generally thought of as AI literacy on the one hand, and the situated literacies that are formed by technological mediation of actual AI technologies on the other. In this regard, the entoptic metaphor can serve as one example for RtD engaging this gap fruitfully, deriving implications and conceptual vocabularies that could not have preceded the actual engagement of designers or other stakeholders with AI technologies. As shown in the secondary users' reflections, developing a practice with and a related situated literacy for the Entoptic Field Camera proved highly individualized and contextualized; shaping subjective choices in tandem with the processing of inputs and outputs. In this light, we propose that designing via the entoptic metaphor produces artefacts of another kind than, for instance, the co-design artefacts which have been employed in the XAI or FAccT space (see e.g.~\cite{benjamin_explanation_2022,xie_chexplain_2020,wang_designing_2019}). Accordingly, we ask ourselves what exactly it is about the entoptic metaphor that `works,' and whether this implies a particular role for RtD in this space.

Considering what the Entoptic Field Camera articulates about its underlying AI technologies (i.e., the BigBiGAN and HiFill models), the entoptic metaphor is arguably not bringing anything `hidden' to light such as Crawford and Joler’s \textit{Anatomy of an AI System} ~\cite{crawford_anatomy_2018}---there is no direct revelatory or pedagogical effect. However, the entoptic metaphor nonetheless produced a multiplicity of perspectives which linked technical aspects to particular concerns, as evidenced by our findings. We argue that exactly this is a valuable programmatic implication: rather than assuming that technical or corporate opacity can be overcome (see also ~\cite{benjamin_machine_2021}), the entoptic metaphor produces a \textit{secondary} opacity that is detached from the original purpose of the AI technology at hand---e.g., `improving' images through GANs. Accordingly, designing alternative, symbolic forms of opacity may be a particularly useful way for RtD to expand beyond the recognized forms of opacity (i.e., technical) of AI technologies. Notably, this reflects Ghajargar and Bardzell's work which frames RtD as a type of synthesis that bridges ``conventional'' and ``evocative'' forms~\cite{ghajargar_synthesis_2021}. In the case of AI technologies, we suggest that the entoptic metaphor is synthetic in a similar way yet on the conceptual level of opacity. As an ``analogical friction''~\cite{pierce_tension_2021} to the \textit{conventional} use of GANs in photography, the Entoptic Field Camera can be interpreted as an example of RtD in which insights can be gathered on a conventional form of opacity (i.e., technical) through engagement with an evocative one (i.e., metaphorical). 

Metaphor-driven research products such as the Entoptic Field Camera probe for phenomena in worlds made legible by AI technologies. RtD's privilege to engage in `practical experiments' allows such probing into a ``pre-language, pre-predicative, a pre-discursive''~\cite{waldenfels_time_2000} space of propositions---the latter holding latent implications for design practice, theory and research or other forms of knowledge making. In effect, by offering a shortcut across the tensions of AI materiality and literacy, RtD can therefore avoid what anthropologist Lee has referred to as a persistent ``epistemic trap''~\cite{lee_enacting_2021} in discourses around AI technologies, and particularly the idea of literacy through transparency. Lee, drawing from Callon~\cite{callon_techno-economic_1990}, argues that studies need to progress beyond seeing AI technologies as ``punctualized;'' i.e. as finite objects that need to be explained or made transparent. This, we argue, is a key opening for RtD with regards to AI literacy. Across many fields of study (e.g., HCI, FAccT, XAI, anthropology, sociology), it is a systemic problem that researching, critiquing or designing with AI technologies generally relies on either a priori (e.g., the arrangement of software components, the provenance of data) or post-hoc (e.g., outputs, socio-technical consequences) objects and events. The entoptic metaphor, then, shows how RtD researchers can instead zero in on the continuous interplay of either. This allows researchers to move from the epistemic trap of what \textit{is} there to how \textit{there} is being made---or, as we have stressed, move from what AI technologies \textit{are} to what they \textit{do}.
\subsection{Entoptic Affordances and 21st Century Design}
Lastly, our engagement with the Entoptic Field Camera has opened up a promising research trajectory on the materiality of AI technologies, design affordances, and the conditions of 21st century design; which we discuss briefly in the following. This contribution is informed by Benjamin's experience in designing the Entoptic Field Camera, Biggs's mixed-media experimentation, as well as Pierce's probing of recursive input-output pairings. We propose that all these, as well as the implications derived from our study, are indicative of a new ``formative''~\cite{cassirer_language_1953} structure of design affordances shaped by 21st century technologies such as GANs.
The notion that forms of expression are related to technological development, and particularly optical techniques (cf. ~\cite{kittler_optical_2010}), is nothing new. Indeed, this has been demonstrated in studies as early as Panofsky's seminal 1926 analysis on the differences between Ancient Greek ``aggregate space,'' where sculptures mapped to `natural' perception and ornamentation to simple layering of shapes, and post-Renaissance ``systematic space''~\cite{panofsky_perspective_1991}, where linear perspective allowed for the geometrization of the entire sensible world. 
More radically, a \textit{co-evolution} of forms of expression and technological development has been substantiated by philosophical anthropologist L{\"o}ffler, who draws on empirical work to show how technologies both seed as well as draw from particular ``affordant ontologies''~\cite{loffler_distributing_2018,loffler_generative_2019}. The latter can be described as historically evolving material assemblages of human and non-human actors which shape how the world is accessible to human intervention; and, importantly, thereby shaping both the concrete forms of expression (e.g., linear perspective paintings) as well as the governing metaphysical principle (e.g., geometric `rationality'). 

In this light, we call for a more comprehensive theoretical and designerly engagement with the forms of expression afforded to design by contemporary AI technologies (e.g., deep neural networks, GANs, prompt-based image synthesis, text generation, etc.). We do so because a more exhaustive reckoning with the formative structure of 21st century design is not only an intriguing theoretical endeavor, but also critically important for design: The early 21st century is a time in which the active pursuit of alternative forms of production, representation, practice and subjectivity are of existential importance. Accordingly, if design can gain an active foothold on its contemporary formative structure (i.e., how design affordances are made available in light of technological development), it is more likely to distinguish itself from---or at least, gesture beyond---the catastrophic tendencies of the forms expressing and perpetuating extractive capitalism. We see the presented work and implications for design by the entoptic metaphor in exactly this light; allowing for first thoughts on the matter which we sketch in the following.

Designing as well as using the Entoptic Field Camera involved conceiving and putting into action forms that in themselves were absent from intentional design practice---in other words, the actual entoptic image could not be designed to the same degree as, for instance, a graphic user interface or chair can be designed. As AI technologies rapidly enter designers' toolkits, we therefore propose that there is a particular \textit{material absence} which characterizes the novel entoptic affordances for design. While absence in a direct sense (e.g., as negative space, darkness, silence) is a relatively well-known or at least intuitively engaged design material, our engagement with the entoptic metaphor also attests that the absence `encountered' in AI technologies is different to to this `first-order' absence---it is neither ``ready-at-hand'' nor ``present-to-hand''~\cite{heidegger_being_2010} as an intuitive or deliberate way to make other forms stand out through, e.g., visual gaps or pauses in performance. Our work with the entoptic metaphor, we argue, is an example for RtD that leverages the contemporary formative structure by designing with the `absent' technological affordance of the metaphorical entoptic field. However, it should be noted that this does not in itself articulate what the formative structure \textit{is}, or what e.g. particular ethico-political dimension are that it is associated with. We propose that a more exhaustive theoretical engagement, which compares the evolution of design affordances---and particularly, forms of absence---across various historical formative structures (e.g., Western linear perspective or Japanese \textit{ma}), may be required. In this regard, we propose that metaphor-driven RtD such as the Entoptic Field Camera can critically support, scaffold and even lead such theoretical work.
\subsection{Limitations and Future Work}
Given the experimental approach of RtD projects, there are limitations to the presented work. First, the selection of participants in the field study introduces significant bias due to their expertise and further involvement as co-authors. Arguably, the core motivation for developing the entoptic metaphor was to unfold ongoing reality-shaping influences of AI technologies in everyday, consumer-grade technological products---suggesting that the involvement of a more diverse set of stakeholders, with more everyday connections to AI technologies or the systems they are embedded in, could have been pursued. However, we argue that this initial restriction has led to design opportunities and implications for design methodologies that may allow for a subsequent broadening of the people involved.
Second, and relatedly, it may be asked who we envision gets to make entoptic media? Power imbalances are particularly prevalent and intricate surrounding AI technologies, while our implications are thus far `only' oriented at future designers. Accordingly, we argue that further work is needed to ensure metaphor-driven RtD centered on AI technologies should participatory methods. Dove and Fayard are exemplary in this regard, bringing a basic metaphoric concept and letting participants iterate around it~\cite{dove_monsters_2020}.

Lastly, while we suggest that critical work can be done via the entoptic metaphor, it is unclear how far this can be pursued. Particularly difficulty resides around the question of datasets: if entoptic media are essentially metaphor-driven RtD experimentation with actual AI technologies, then arguably they may arrive `late to the scene;' i.e. after significant political and ethical dimensions around dataset curation and gathering have already played out. Again, however, we argue that the role of such prototypes needs to be borne in mind: these are not tools for auditing purposes, but rather artefacts that prompt articulations of the effects of AI technologies. Nonetheless, the possibility of the entoptic metaphor in itself to inform actual \textit{technical} designs of, e.g., AI pipelines in future work remains intriguing: if we pay attention to the concept of material interplay, do new opportunities arise in which, for instance, users could flag or contest overt transformations stemming from specific cases of interplay? As such, we argue that this project has laid the groundwork for rich and manifold future explorations.

\section{Conclusion}
In this paper, we have presented a metaphor-driven RtD project: the Entoptic Field Camera. It was motivated by our intuition that enthusiasm over current advances in prompt-based image synthesis models may overshadow the often subtle ways that AI technologies shape experiences of reality. Through the development of the entoptic metaphor, we introduced an analogy that thematized the latter: just as the physiological interplay of eye and brain lead to involuntary perceptions (of, e.g., floaters), so the material interplay of AI technologies' components (i.e., models, datasets, algorithms) can shape how realities are experienced. We materialized this metaphor in the form of a research product, the Entoptic Field Camera, a web application with which users take an image that is subsequently processed by GAN techniques. Through a design process informed by autobiographical design and a subsequent field study among co-authors, we found that the Entoptic Field Camera prompted specific forms of practice and led to the articulation of particular situated literacies. Taken together, these provided  design implications for critical, reflective, more-than-human, sustainable and ludic design to engage with AI technologies. Due to the breadth of implications, we further reflected on programmatic implications, where we argued that RtD may contribute to discourses around AI literacy by materializing prompts for situated articulations that reflect actual AI technologies without relying on conventional criteria of technical literacy or opacity. Lastly, we reflected on whether the entoptic metaphor, as an instance of designerly ways of engaging AI technologies, may indicate specificities of 21st century design, and sketched out an initial research trajectory. In conclusion, we argue that metaphor-driven RtD projects such as the one presented hold tremendous potential for the HCI community to further probe at the ``curvatures''~\cite{ihde_technology_1990} of the coming 21st century lifeworld.

\begin{acks}
We wish to thank our reviewers for their constructive feedback, Tim Korjakow for making the entoptic field addressable, and Ian Forrester for providing a venue for real-world encounters. This work is supported by UK Research and Innovation (grant MR/T019220/1, ``Design Research Works'') and the National Science Foundation, grants \#2142795 and \#2230825.
\end{acks}

\bibliographystyle{ACM-Reference-Format}
\bibliography{main}


\end{document}